\documentclass[
reprint, 
superscriptaddress,
nofootinbib,
 amsmath,amssymb,
 aps,
 prresearch,
 physrev,
dvipdfmx, 
]{revtex4-2}

\usepackage{graphicx}
\usepackage{dcolumn}
\usepackage{bm}

\usepackage{color}

\newcommand{\RED}[1]{{{\color{red}#1}}}
\newcommand{\GREEN}[1]{{{\color{green}}}}

\usepackage{uline--} 
\usepackage{here}

\usepackage[symbol]{footmisc}

\begin{document}

\preprint{APS/123-QED}

\title{Inference of noise intensity and phase response from noisy synchronous oscillators}

\author{Hisa-Aki Tanaka}  \email{htanaka@uec.ac.jp}
 \affiliation{The University of Electro-Communications, Tokyo 182-8585, Japan}
\author{Somei Suga}
 \affiliation{The University of Electro-Communications, Tokyo 182-8585, Japan}
\author{Akira Keida}
 \affiliation{The University of Electro-Communications, Tokyo 182-8585, Japan}
\author{Hiroya Nakao}
 \affiliation{Tokyo Institute of Technology, Tokyo 152-8552, Japan} 
\author{Yutaka Jitsumatsu} 
 \affiliation{Tokyo Institute of Technology, Tokyo 152-8552, Japan}
\author{Istv\'{a}n Z. Kiss}
 \affiliation{Department of Chemistry, Saint Louis University -- St. Louis, MO 63103, USA}



%

\date{\today}

\begin{abstract}
Numerous biological and microscale systems exhibit synchronization in noisy environments.
The theory of such noisy oscillators and their synchronization has been developed and experimentally demonstrated, but inferring the noise intensity and phase response is not always straightforward.
In this study, we propose a useful formula that enables us to infer the noise intensity and phase response of a noisy oscillator synchronized with periodic external forcing.
Through asymptotic approximations for small noise, we show that noisy synchronous oscillators satisfy a simple relationship among the noise intensity and measurable quantities, i.e., the stationary distribution of the oscillation phase and stationary probability current obtained as the average phase velocity, which is verified through systematic numerical analysis.
The proposed formula facilitates a unified analysis and design of synchronous oscillators in weakly noisy environments.
\end{abstract}


\keywords{subject area : Biological Physics, Computational Physics}
\maketitle

\section{\label{sec:introduction}Introduction}

Noisy oscillators are ubiquitous in biological and micro/nanoscale systems \cite{Wilson2022,Wilson2015,Ota2009,Ermentrout2007,Galan2005,Teramae2004,Ermentrout2010,Pikovsky2001,Toyabe,Kitaya2022,Blickle2012,Izumida2016,Toyabe2020,Izumida2020,Knunz2010, Sadeghpour2013,Noh2020,Lifshitz2021}.
Examples have been reported in various branches of physics, including biological physics \cite{Monga2019,Bodenschatz2016}, computational physics \cite{Kitaya2022}, electronics \cite{Viterbi1966,Udo2012,Ciliberto2017}, and nanophysics \cite{Knunz2010,Sadeghpour2013,Noh2020,Lifshitz2021}, where synchronization of noisy oscillators have been studied.
Thus, a method for inferring the noise intensity and phase response of such oscillators based on easily measurable observables is desirable.
In this study, we consider this problem for weakly noisy synchronous oscillators.

Previous studies have investigated the phase dynamics of noisy oscillators and refined the definitions of the asymptotic phase \cite{Schwabedal2013,Thomas2014,Cerecera2023,Kato2019}.
In particular, theoretical \cite{Ermentrout2007,Ota2009} and experimental \cite{Galan2005,Imai2017} studies have been conducted to infer the key information intrinsic to the oscillator, namely the phase response property characterized by the phase sensitivity function (also known as infinitesimal phase resetting curve) \cite{Wilson2015,Ermentrout2010,Wilson2022,Ermentrout2007,Ota2009,Galan2005, Imai2017,Stankovski2017,Pikovsky2001,Kuramoto}, for real noisy neurons \cite{Galan2005} and van der Pol--type electrical circuits \cite{Imai2017}.
In addition to the phase sensitivity function, the phase coupling function defined in Sec.~\ref{sec:2}, has been experimentally measured \cite{Miyazaki2006,Stankovski2017}.

Despite these theoretical and experimental advances, how to evaluate the noise intensity and phase sensitivity function (PSF) of a noisy synchronous oscillator remains yet to be clarified.
It has recently been reported that the effective noise intensity can be inferred in amoeboid cells near the onset of spontaneous oscillations from the probability density function (PDF) of the squared amplitude of oscillations observed experimentally.
In the present study, we provide a general framework for inferring both the effective noise intensity and the PSF of the oscillator at the same time.
Our framework uses the coherence measure of oscillations that can be obtained from the PDF of the oscillator phase, and it is applicable to experimental systems in weakly noisy environments.

The rest of this paper is organized as follows.
In Sec.~\ref{sec:2}, we introduce phase equations \cite{Ermentrout2010,Pikovsky2001,Kuramoto} for noisy synchronous oscillators and characterize experimentally measurable quantities.
In Sec.~\ref{sec:3}, we introduce an observable, i.e., a coherence measure of oscillators under noisy synchronization.
In Sec.~\ref{sec:beta-and-Z}, through asymptotic approximations for small noise, we derive simple relations that enable us to infer the effective noise intensity (\textit{Result 1} and \textit{Result 2}) and the PSF (\textit{Result 3}) of the oscillator.
In Sec.~\ref{sec:discussions}, we discuss possible applications of the proposed framework in several branches of physics.
Appendices provide details of asymptotic approximations and relations among the noise intensity, coherence measure, and various measurable quantities.
\begin{figure}[h]
\includegraphics[scale=0.35]{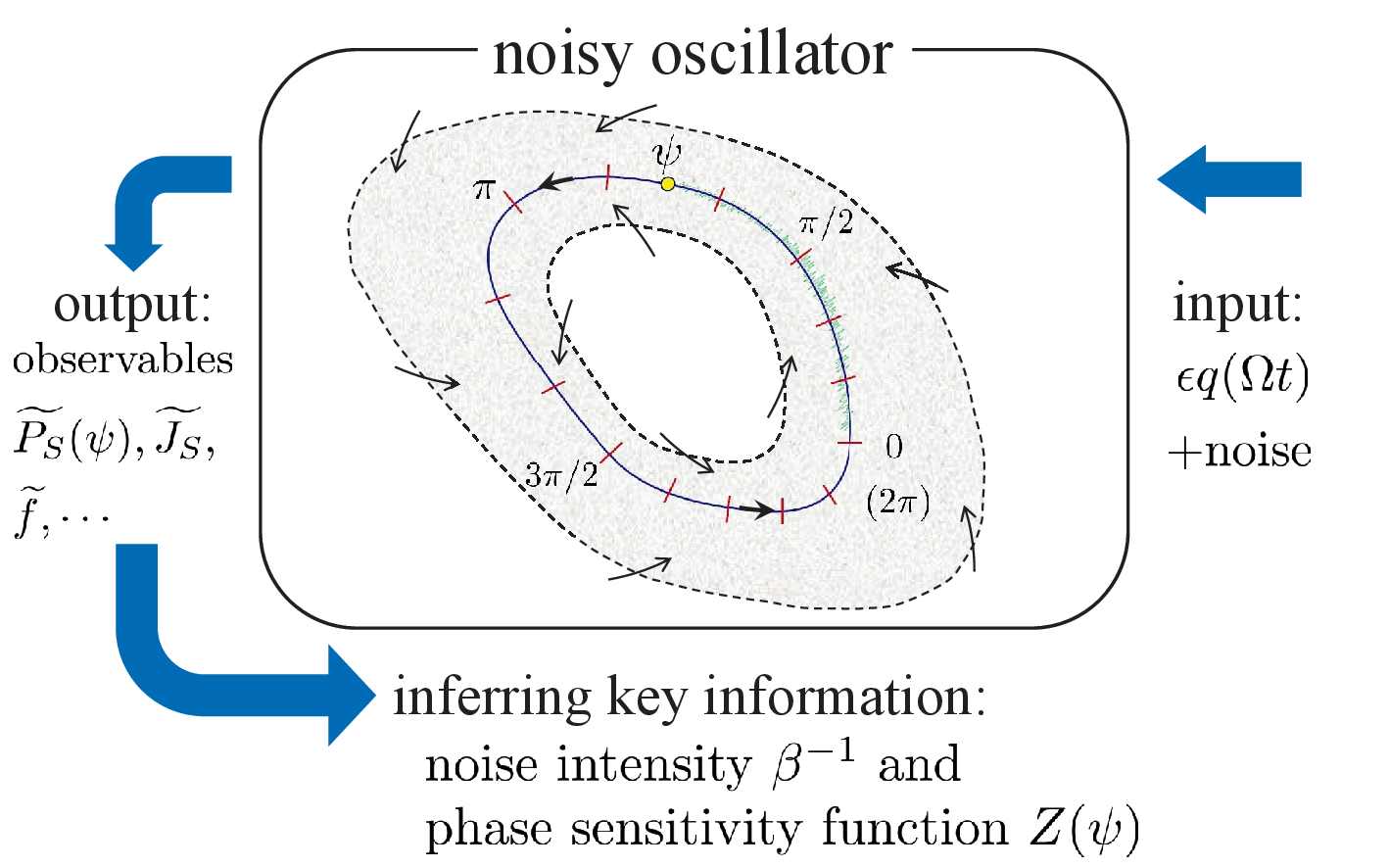}
\caption{Proposed framework for inferring the noise intensity and phase sensitivity function, $\beta^{-1}$ and $Z$.
Through phase reduction in terms of the oscillation phase $\psi$, an oscillator subjected to an external forcing $\epsilon q$ and noise $\sqrt{\epsilon}\eta$ obeys a Langevin equation (Eq.~\eqref{eq:0209_dot_psi_5}).
From the measured stationary probability density function $\widetilde{P_S}$, constant probability current $\widetilde{J_S}$ and potential slope $\widetilde{f}$, etc., we infer $\beta^{-1}$ and $Z$.
}
\label{fig_noisyOscillator}
\end{figure}

\section{\label{sec:2}{Noisy synchronous oscillators}}

In this section, we introduce the essential building blocks of the proposed theory for noisy synchronous oscillators, namely phase equation \cite{Ermentrout2010,Pikovsky2001,Kuramoto} and several experimentally measurable quantities \cite{Udo2012,Ciliberto2017}.

\subsection{Phase equation}

First, we explain the phase equation for weakly perturbed limit-cycle oscillators used in this study.
Under a weak periodic external forcing and weak additive environmental noise, the dynamics of a limit-cycle oscillator can be reduced to the following phase equation by using phase reduction theory \cite{Kuramoto}:
\begin{align}
\dot\phi = \omega + Z(\phi) \left[\epsilon q(\Omega t) + \sqrt{\epsilon} \xi(t)\right],
\label{eq:0209_dot_phi_1}
\end{align}
where $\phi$, $\omega$, and $Z(\phi)$ are the oscillator's phase, natural frequency, and PSF \cite{Kuramoto}, respectively, $q(\Omega t)$ is the periodic external forcing of frequency $\Omega$, $\xi(t)$ is the environmental noise, and $\epsilon~(>0)$ is a small parameter that characterizes the magnitude of the weak external forcing and small noise.

Equation \eqref{eq:0209_dot_phi_1} can also be used to model an ensemble of uncoupled limit-cycle oscillators driven by a common forcing (i.e., all oscillators are subjected to the same forcing).
Coherent oscillations in such an ensemble are relevant, e.g., in combining the output power of semiconductor laser arrays, understanding circadian oscillations in cells, and designing chronotherapy for acute leukemia (cf. \cite[Chap. 3, pp. 98--101]{Pikovsky2001}).
In two mutually coupled noisy oscillators, such as two mutually coupled electrical oscillators or two noisy pacemaking neurons, their phase difference also satisfies an equation similar to Eq.~(\ref{eq:0209_dot_phi_1}) (cf.~\cite[Chap. 9, pp. 245--246]{Pikovsky2001}).

Although non-Gaussian colored noise is an interesting and important class of problems (e.g., \cite{Goldobin2010}), as the simplest setting, we assume $\xi$ in Eq.~\eqref{eq:0209_dot_phi_1} to be white Gaussian noise with zero mean and the correlation function given by $\langle\xi(t_0)\xi(t_0+t)\rangle=2\alpha^{-1}\delta(t)$, where $\alpha^{-1}$ is the intensity of $\xi$, $\delta(t)$ is Dirac's delta, and $\langle\cdot\rangle$ denotes the average over noise realizations.
The magnitude $\epsilon$ of the forcing and noise is assumed to be sufficiently small such that a phase-only description of the oscillator, namely Eq.~(\ref{eq:0209_dot_phi_1}), holds \cite{Pikovsky2001,Kuramoto}.
However, in practice, Eq.~(\ref{eq:0209_dot_phi_1}) (and the time-averaged Eq.~(\ref{eq:0209_dot_psi_5})) can serve also as an adequate qualitative model even for experimental systems with a moderate magnitude of $\epsilon$ (cf.~\cite{Wilson2015,Imai2017, Galan2005,Stankovski2017,Miyazaki2006,Harada2010,Zlotnik2013}).

\subsection{Derivation of the averaged phase equation}

Next, we derive the averaged phase equation by using the averaging method.
As discussed in \cite{Kawamura2007,schmidt2014}, we can selectively average the right-hand side (RHS) of Eq.~(\ref{eq:0209_dot_phi_1}), including the noise term, via the near-identity transformation (cf.~\cite[p. 80, exercise 3.2.6]{Strogatz1994})\footnote[2]{The near-identity transformation is a method that transforms the original phase equation into a simpler phase equation by introducing a new, slightly deformed phase variable as $\phi=\phi^{\prime} + \epsilon h(\phi^{\prime})$ and rewriting the original phase equation (for $\phi$) using the new phase variable $\phi^{\prime}$, which typically yields the same result as averaging the terms on the RHS of the original phase equation.
Here, we selectively averaged the noise term in Eq.~\eqref{eq:0209_dot_phi_1} to derive Eq.~\eqref{eq:0209_dot_phi_2}.
A detailed derivation can be found in \cite{Kawamura2007,schmidt2014} and \cite{Kato2019}.
Note that we use the same symbol $\phi$ for the phase variable after the averaging, as the difference from the original phase is of $O(\epsilon)$ and small.},
yielding the approximate phase equation 
\begin{align}
\dot\phi = \omega + \epsilon Z(\phi) q(\Omega t) + \sqrt{\epsilon} \eta(t),
\label{eq:0209_dot_phi_2}
\end{align}
where $\eta(t)$ represents zero-mean white Gaussian noise whose correlation function is given by $\langle\eta(t_0)\eta(t_0+t)\rangle = 2 \beta^{-1}\delta(t)$;
the effective noise intensity $\beta^{-1}$ is given by
\begin{align}
\beta^{-1} = \alpha^{-1} \frac{1}{2\pi} \int_0^{2\pi} d\psi Z(\psi)^2.
\end{align}

We further assume that the frequency mismatch between $\omega$ and $\Omega$ is sufficiently small and of $O(\epsilon)$, and denote it as $\epsilon d = \omega - \Omega$.
By introducing a relative phase of the oscillator with respect to the periodic external forcing, $\psi = \phi - \Omega t$, Eq.~(\ref{eq:0209_dot_phi_2}) is rewritten as
\begin{align}
\dot\psi = \epsilon d + \epsilon Z(\psi + \Omega t) q(\Omega t) + \sqrt{\epsilon} \eta(t).
\label{eq:0213_dot_psi_4}
\end{align}
As discussed in previous studies (e.g., \cite{Kuramoto}), we can further average the coupling term on the RHS of Eq.~(\ref{eq:0213_dot_psi_4}) and obtain the phase equation \cite{Ermentrout2010,Pikovsky2001,Stankovski2017,Kuramoto}
\begin{align}
\dot\psi = \epsilon [d + H(\psi)] + \sqrt{\epsilon}\eta(t),
\label{eq:0209_dot_psi_5}
\end{align}
where the phase coupling function $H(\psi)$ \cite{Kuramoto} is given by
\begin{align}
H(\psi) = \frac{1}{2\pi} \int_0^{2\pi} d\theta Z(\theta+\psi)q(\theta).
\label{define_Gamma}
\end{align}
The Langevin equation \eqref{eq:0209_dot_psi_5} is equivalent to an Ito stochastic differential equation (SDE) of the form
\begin{align}
d\psi & = \epsilon [d + H(\psi)]dt + \sqrt{\epsilon}\sqrt{\beta^{-1}}dW(t),
\label{eq:0428_dpsi/dtau}
\end{align}
where $W(t)$ represents a Wiener process. 
We denote the constant part of $d + H(\psi)$ by $f$ and define a tilted potential $V$ that satisfies $-V^{\prime}(\psi) = d + H(\psi)$:
\begin{align}
V(\psi)=
-\int^{\psi}\mathit{d}\psi^{\prime}[d+H(\psi^{\prime})]
=\overline{V}(\psi) - f \psi, 
\label{vpsi_fpsi}
\end{align}
where we defined a purely periodic function $\overline{V}(\psi)~\left(=\overline{V}(\psi + 2\pi)\right)$.

This SDE \eqref{eq:0428_dpsi/dtau} can be transformed into a Fokker--Planck equation for the PDF $P(\psi,t)$ (abbreviated $P$ below) \cite{Udo2012,Kato2019,Kawamura2007} after setting $\epsilon$ to unity by time-rescaling:
\begin{align}
\frac{\partial}{\partial t}P(\psi,t)
= -\frac{\partial}{\partial \psi} \left\{\left[d + H(\psi)\right]P\right\} + \beta^{-1}\frac{\partial^2}{\partial\psi^2}P.
\label{eq:0428_partian_tPpsit}
\end{align}
For this equation, it is known that the stationary PDF $P_{S}(\psi)$, the constant probability current $J_{S}$ defined as $J_{S} \equiv \left[d + H(\psi) \right] P_S(\psi) - \beta^{-1}P_S^{\prime}(\psi)$, and the normalization constant $C$ have the following expressions \cite{Risken1996, Pikovsky2001}:
\begin{align}
P_{S}(\psi)&=C^{-1}\int_{\psi}^{\psi+2\pi}d\psi^{\prime}e^{\beta \left[V(\psi^{\prime})-V(\psi)\right]},
\label{form_Ps}\\
J_{S}&=\beta^{-1}{C}^{-1}\left(1-e^{-2 \pi f \beta}\right),
\label{form_Js}\\
\text{with}~
C&=\int_0^{2\pi}\int_{\psi}^{\psi + 2\pi} d\psi^{\prime}d\psi e^{\beta [V(\psi^{\prime}) - V(\psi)]}.
\label{form_C}
\end{align}

\subsection{Experimentally measurable quantities}

In this study, we assume that the stationary PDF $P_{S}(\psi)$ and the probability current $J_{S}$ can be measured; $J_S$ is obtained through the time average of the phase velocity $\dot{\psi}$ since $J_{S}=(2\pi)^{-1}\langle\dot{\psi}\rangle$ \cite{Blickle2007a, Blickle2007, Speck2006, Speck2007}.
Reliable data for the measured $P_{S}(\psi)$ and $J_{S}$ have been obtained, for example, for microsystems of colloidal particles \cite{Udo2012, Ciliberto2017, Blickle2007a, Blickle2007, Speck2007} and for certain electronic circuits \cite{Viterbi1966} by experiments and simulations.
In addition to $P_{S}(\psi)$ and $J_{S}$, the potential slope $f$ can also be measured by calculating $f=(2\pi)^{-1}\int^{2\pi}_{0} d\psi J_{S}P_{S}(\psi)^{-1}$, as shown in Eq.~(\ref{define_Q}).
The potential $V(\psi)$ is reconstructed from $\beta$, $J_{S}$, and $P_S(\psi)$ (cf.~\cite{Udo2012, Blickle2007a, Blickle2007, Speck2006, Speck2007}), as shown in Eq.~(\ref{-Vd}).

In what follows, only when it is necessary, we denote the measured (or estimated) value of a physical quantity $\alpha$ as $\widetilde{\alpha}$ to avoid confusions;
for instance, in the numerical verifications described below, $\widetilde{P_S}$ and $\widetilde{J_S}$ are measured through the numerical integration of Eq.~(\ref{eq:0428_dpsi/dtau}), as opposed to the true $P_S$ and $J_S$, respectively.
Figure~\ref{fig_noisyOscillator} shows the proposed framework for inferring $\beta^{-1}$ and $Z$ from the measured observables $\widetilde{P_S}$, $\widetilde{J_S}$, and $\widetilde{f}$.

\section{\label{sec:3}coherence of oscillations}

\subsection{Background and basic properties of the potential}

Here, we define the coherence of noisy oscillators under synchronization with the external forcing using only directly measurable quantities.
Maximization of the coherence of oscillations in noisy oscillators with external forcing was independently addressed in \cite{Pikovsky2015} and \cite{Wilson2015}.
In \cite{Pikovsky2015}, the diffusion constant $D$ was associated with the coherence ($\mathcal{C}_{D}$ defined below) for a noisy oscillator with the potential $V$ defined above and driven by the white Gaussian noise $\eta$ \cite{Bire82}, where smaller value of $D$ implies higher coherence of oscillations.
On the other hand, in \cite{Wilson2015}, the height of the potential barrier is associated with the coherence; a higher potential barrier implies a longer average escape time over the potential barrier, which can be used as another coherence measure ($\mathcal{C}_{S}$ defined below).

\begin{figure}[h]
\begin{center}
\includegraphics[scale=0.75]{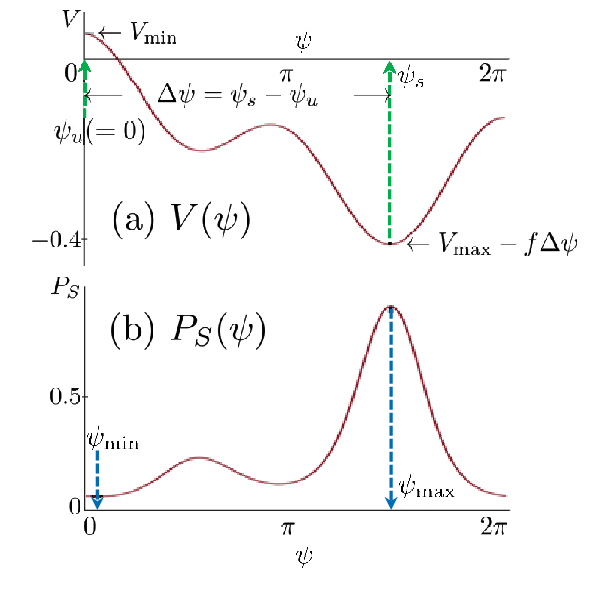}
\caption{Graphs of $V(\psi)$ and $P_S(\psi)$.
(a)~$\psi_u$ and $\psi_s$ are respectively the maxima and minima of $V(\psi)$.
(b)~$\psi_{\max}~(\rightarrow\psi_{s})$ and $\psi_{\min}~(\rightarrow\psi_{u})$ are respectively the maxima and minima of $P_S(\psi)$ (as $\vert f\vert\rightarrow0$ or as $\beta\rightarrow\infty$).
These graphs were obtained for the Example in Eq.~(\ref{Z_theta}) with $r=0.2$, $\beta=10.0$, and $f=0.28\times10^{-1}$.}
\label{fig_potential}
\end{center}
\end{figure}

Regarding these two coherence measures, we have the following basic properties of the potential $V(\psi)$.
Firstly, the maximum and minimum values of $V(\psi)$ for $\psi \in [0,2\pi]$ are assumed to be attained at $\psi = \psi_{u}$ and $\psi_{s}$, respectively, where $-V^{\prime}(\psi_{u,s})=d+H(\psi_{u,s})=0$, as shown in Fig.~\ref{fig_potential}(a).
Here, $\psi_{u}$ and $\psi_{s}$ respectively correspond to the unstable and stable fixed points of the phase equation $\dot{\psi}=d+H(\psi)$ \cite{Pikovsky2001,Kuramoto}, satisfying
\begin{align}
H^{\prime}(\psi_{u}) = -V^{\prime\prime}(\psi_{u}) > 0,~
H^{\prime}(\psi_{s}) = -V^{\prime\prime}(\psi_{s}) < 0.
\label{gammas}
\end{align}

On the other hand, we define the maxima and minima of $P_S(\psi)$ as $\psi=\psi_{\max}$ and $\psi_{\min}$, respectively, which satisfy $P_S^{\prime}(\psi_{\max})=P_S^{\prime}(\psi_{\min})=0$, as shown in Fig.~\ref{fig_potential}(b).
Then, we obtain
\begin{align}
\psi_{\max}\approx\psi_{s},~
\psi_{\min}\approx\psi_{u},
\label{eq:psi_max-min}
\end{align}
asymptotically for $\vert f \vert \ll 1$ or for $\beta \gg 1$, as follows.
First, after integrating the RHS of Eq.~(\ref{eq:0428_partian_tPpsit}) and plugging $P_S^{\prime}(\psi_{\max,\min})=0$ into the result, we obtain
\begin{align}
&\left[d+H(\psi_{\max})\right]P_{S}(\psi_{\max}) \notag \\
&= \left[d+H(\psi_{\min})\right]P_{S}(\psi_{\min})
= J_{S}.
\label{eq:0518_Delta+GammaP_S}
\end{align}
Next, as $\vert f \vert\rightarrow0$ or as $\beta\rightarrow\infty$, $\vert J_{S} \vert\rightarrow0$ follows immediately from Eq.~(\ref{form_Js}) and hence $d + H(\psi_{\max,\min})\rightarrow0$ in Eq.~(\ref{eq:0518_Delta+GammaP_S}), which implies Eq.~(\ref{eq:psi_max-min}).
Here, we note that for all the Examples in Sec.~\ref{sec:beta-and-Z}, $\psi_{\max,\min}\approx\psi_{s,u}$ in Eq.~(\ref{eq:psi_max-min}) can be numerically verified.

\subsection{Two measures of coherence}
With the above basic properties in mind, we consider two measures of coherence, namely $\mathcal{C}_{D}$ and $\mathcal{C}_{S}$, as follows.
First, following \cite{Pikovsky2015}, we obtain $\mathcal{C}_{D}$ from the diffusion constant $D$ as 
\begin{align}
\mathcal{C}_{D} \equiv D_{0}/D,
\end{align}
where $D$ is defined later in Eq.~\eqref{D_D_0} and $D_{0}$ is the bare diffusion constant without the external forcing $q(\Omega t)$;
a lower $D$ implies a higher $\mathcal{C}_{D}$.
Next, we introduce another measure, namely $\mathcal{C}_S$, as 
\begin{align}
\mathcal{C}_{S} \equiv P_{S}(\psi_{\max})/P_{S}(\psi_{\min}),
\end{align}
where $\psi_{\max}$ and $\psi_{\min}$ are the maxima and minima of the stationary PDF $P_{S}(\psi)$ in Eq.~(\ref{form_Ps}), respectively.

To gain insights for $\mathcal{C}_{D}$ and $\mathcal{C}_{S}$, we start from a physical interpretation of them.
Here, we focus on the case of $f=0$ without loss of generality.
First, after eliminating $C^{-1} \int_{\psi}^{\psi+2\pi} d{\psi}^{\prime}e^{\beta V({\psi}^{\prime})}$ by dividing $P_S(\psi_{\max})$ by $P_S(\psi_{\min})$ in Eq.~(\ref{form_Ps}), the following equation holds for any $\beta$:
\begin{align} 
\mathcal{C}_{S}
= P_{S}(\psi_{\max})/P_{S}(\psi_{\min})
= e^{\beta \Delta\overline{V}},
\label{eq:C_S_1003}
\end{align}
where $\Delta\overline{V}$ represents the height of the potential barrier in $\overline{V}(\psi)$ ($= V_{\min}-V_{\max}$, defined in Eq.~(\ref{eq:0515_V})).
This characterizes the local information of the potential $\overline{V}(\psi)$ at the two points $\psi=\psi_{\max}$ and $\psi=\psi_{\min}$ as the shown in Fig.~\ref{fig_potential};
a higher potential barrier $\Delta\overline{V}$ implies a higher $\mathcal{C}_{S}~\left(=e^{\beta\Delta\overline{V}}\right)$ (cf. \cite[Sec.~2]{Wilson2015}).

In contrast, $\mathcal{C}_{D}$ has more detailed information on $V$ since
\begin{align}
\mathcal{C}_{D}\approx 2\pi\beta^{-1}\left[-H^{\prime}(\psi_{\max})H^{\prime}(\psi_{\min})\right]^{-\frac{1}{2}} e^{\beta \Delta\overline{V}}
\label{eq:C_D_0729}
\end{align}
for sufficiently large $\beta$, which involves the curvatures $V^{\prime\prime}~\left(=-H^{\prime}\right)$ at the maximum and minimum of $V$.
The derivation of Eq.~(\ref{eq:C_D_0729}) is given in Appendix \ref{sec:appendix_B}.

\subsection{Asymptotic approximations of the two measures of coherence}

For the two measures $\mathcal{C}_{S}$ and $\mathcal{C}_{D}$ introduced above, asymptotic approximations are valid for the following limiting cases;
\textbf{Case (i)}:
$\beta\rightarrow\infty$ and $\vert f\vert\rightarrow0$
\cite[Chap. 8, Sec. 10, Eq. (10.10)]{R.Wong1989}, and \textbf{Case (ii)}:
$\beta\rightarrow\infty$ and $\vert f\vert\rightarrow\infty$ \cite[Chap. 2, Sec. 1, Theorem 1, Eq. (1.12)]{R.Wong1989}.
We note that in all the examples given in this study, the asymptotic approximations employed here turn out to be valid for sufficiently large $\beta$ with sufficiently small $\vert f\vert$ or sufficiently large $\vert f\vert$.

First, we consider the \textbf{Case (i)}.
The \textbf{Case (ii)} will be considered in Sec.~\ref{sec:beta-and-Z}.
Here, we notice that $\psi_{u}\rightarrow\psi_{\min}$, $\psi_{s}\rightarrow\psi_{\max}$ in this limit, and $\psi_{\min}$ and $\psi_{\max}$ can be measured from the observable $\widetilde{P_S}$, which implies that $\mathcal{C}_{S}$ can be measured.
For this $\mathcal{C}_S~\left(=P_S(\psi_{\max})/P_S(\psi_{\min})\right)$, from the above definition of the tilted potential $V(\psi)$, shown in Fig.~\ref{fig_potential}(a), by setting the origin of $V(\psi)$ at $\psi = \psi_{u}$, we have the following relationships for sufficiently small $f~(>0)$:
\begin{align}
V(\psi_u)&=V_{\min} \approx V(\psi_{\min}), \notag \\
V(\psi_s)&=V_{\max}-f \Delta\psi \approx V(\psi_{\max}), \notag \\
\Delta V &\equiv V(\psi_u)-V(\psi_s)=V_{\min}-V_{\max} + f \Delta\psi \notag \\
&=\Delta\overline{V}+f\Delta\psi,
\label{eq:0515_V}
\end{align}
where ${\Delta}{\psi} \equiv \psi_{s}-\psi_{u} \approx \psi_{\max}- \psi_{\min}$, and $\Delta\overline{V} \equiv V_{\min}-V_{\max}$.
This definition of $V_{\max}~(<V_{\min})$ looks counterintuitive;
however, it is required for consistency with $P_{S}(\psi_{\max})~(>P_{S}(\psi_{\min}))$.
Then, from the definition of $\mathcal{C}_{S}$ in Eq.~\eqref{eq:C_S_1003}, the following relation is obtained for sufficiently large $\beta$:
\begin{align}
\mathcal{C}_{S}
\approx\frac{2 e^{\beta [V(\psi_{\min})-V(\psi_{\max})]}}{1+e^{\beta 2\pi \vert f \vert}}
\approx\frac{2e^{\beta\Delta V}}{1+e^{\beta 2\pi\vert f \vert}},
\label{eq:0515_C_S}
\end{align}
where $V(\psi_{\min}) - V(\psi_{\max}) \approx \Delta V~(= V(\psi_u)-V(\psi_s))$, as obtained from Eq.~(\ref{eq:psi_max-min}).
We note that relationships similar to those in Eq.~(\ref{eq:0515_V}) also hold for $f<0$ and the same result as that in Eq.~(\ref{eq:0515_C_S}) follows.
The derivation of Eq.~(\ref{eq:0515_C_S}) simply follows from Laplace's formula \cite[Eq.~(1.6)]{R.Wong1989}.

In contrast, for $\mathcal{C}_{D}$, the diffusion constant $D$ is explicitly given by \cite{P.Reimann2001, Hayashi2004, Sasaki2005}
\begin{align}
D&=D_{0}\int_{\psi_{0}}^{\psi_{0}+2\pi} \frac{d\psi}{2\pi} I_{+}(\psi)^{2}I_{-}(\psi)/\left[\int_{\psi_{0}}^{\psi_{0}+2\pi} \frac{d\psi}{2\pi} I_{+}(\psi)\right]^{3} \notag \\
&=D_0\left< I_{+}(\psi)^{2}I_{-}(\psi)\right>_{\psi}/\langle I_{+}(\psi)\rangle_{\psi}^{3},
\label{D_D_0}
\end{align}
where $\psi_0$ is an arbitrary reference phase, $\langle\cdot\rangle_{\psi}\equiv(2\pi)^{-1}\int_{0}^{2\pi}\cdot d\psi$, and
\begin{align}
I_{\pm}(\psi)
\equiv\left< e^{\left[ \pm \beta V(\psi) \mp \beta V(\psi \mp \psi^{\prime}) \right]} \right>_{\psi^{\prime}}.
\label{eq:0612_I_pm(psi)}
\end{align}
It should be noted that Eqs.~(\ref{D_D_0}) and (\ref{eq:0612_I_pm(psi)}) follow the notation in \cite{Sasaki2005}.
From Eq.~(\ref{D_D_0}), the following asymptotic approximation holds for sufficiently large $\beta$:
\begin{align}
\mathcal{C}_{D}
\approx 2\pi\beta^{-1}\left[-H^{\prime}(\psi_{\max})H^{\prime}(\psi_{\min})\right]^{-\frac{1}{2}}e^{\beta\Delta V}
\label{eq:0515_C_D}
\end{align}
with $\Delta V = \Delta\overline{V}+f\Delta\psi$, which is derived through Laplace's approximation of the Laplace-type double integral \cite[Chap. 8, Sec. 10, Eq. (10.10)]{R.Wong1989}.
Note that Eq.~(\ref{eq:C_D_0729}) is obtained by setting $f=0$ in Eq.~(\ref{eq:0515_C_D}).
For detailed derivations of Eqs.~(\ref{eq:0515_C_S}) and (\ref{eq:0515_C_D}), see Appendix~\ref{sec:appendix_A} and Appendix~\ref{sec:appendix_B}, respectively.

Thus, we arrive at the following relationship between $\mathcal{C}_{S}$ and $\mathcal{C}_{D}$:
\begin{align}
\beta^{-1}\ln\mathcal{C}_S - \beta^{-1}\ln\mathcal{C}_D
\rightarrow-4\pi\vert f \vert& \notag \\
(\beta\rightarrow\infty,~\vert f \vert\rightarrow0)&.
\label{relationship_CS_CD}
\end{align}
For a detailed derivation of Eq.~\eqref{relationship_CS_CD}, see Appendix \ref{sec:appendix_C_0716}.

In the context of maximizing the coherence of synchronous oscillators~\cite{Pikovsky2015}, for a given $\beta$ and $f$, $\beta^{-1}\ln\mathcal{C}_S$ and $\beta^{-1}\ln\mathcal{C}_D$ can be maximized by designing the same optimal forcing $q(\Omega t)$, subject to a certain constraint (e.g., a given constant power $\Vert q \Vert_2$ as in \cite{Pikovsky2015};
see \cite{Pikovsky2015} for more details regarding the optimal forcing).
This implies the equivalence of $\mathcal{C}_{S}$ and $\mathcal{C}_{D}$ in maximizing the coherence of weakly noisy oscillations for $\beta \to \infty$ and $\vert f \vert \to 0$.

\section{\label{sec:beta-and-Z}Inference of noise intensity and phase sensitivity function}

\subsection{Outline of the approach and its outcomes}

The main result of this study is that we can infer key information on noisy synchronous oscillators from measurable quantities.
That is, we can infer the noise intensity $\beta^{-1}$ and the PSF $Z$ from the observables including the coherence $\mathcal{C}_S$.

The approach for inferring $\beta^{-1}$ and $Z$ is based on the Langevin dynamics, described by Eq.~(\ref{eq:0209_dot_psi_5}), for noisy limit-cycle oscillators \cite{Pikovsky2015} and the asymptotic approximations of $\mathcal{C}_{S}$ and other related integrals.
Combining the Langevin dynamics and asymptotic approximation can provide new analytical insights into a noisy synchronous oscillator, as presented in \textit{Results 1}, \textit{2}, and \textit{3} below, which provide methods for inferring $\beta^{-1}$ for sufficiently small $\vert f \vert$ or for sufficiently large $\vert f \vert$, and for estimating $Z$ from the obtained $\beta^{-1}$, respectively.
Each of these three results is numerically verified with a simple but nontrivial example.

We note that the above asymptotic approximations (Laplace's approximation) that we use in inferring $\beta$ are known to be practically valid even for an intermediate value of $\beta$.
The following examples in the numerical verifications for \textit{Results 1}, \textit{2}, and \textit{3} show that our framework can be applied to the cases with moderate values of $\beta$ (beyond the weak noise limit $\beta\rightarrow\infty$).

\subsection{Inference of noise intensity from coherence measures}

As mentioned before (see Eqs.~(\ref{eq:0515_V}) and (\ref{eq:0515_C_S})), $\beta$ and $\Delta\overline{V}$ are related to the observables $\mathcal{C}_{S}$, $J_{S}$, $f$, and $\Delta \psi$.
Therefore, we can obtain $\beta$ and $\Delta \overline{V}$ by solving Eq.~(\ref{eq:0515_C_S}) for these measurable quantities.
This is the first main result in this study and can be summarized as follows:

\textit{\textbf{Result 1}}.
Suppose that $\beta$ and $\Delta\overline{V}$ are unknown and $\mathcal{C}_{S}$, $J_{S}$, $f$, and $\Delta\psi$ can be measured;
$\Delta\overline{V}$ represents the depth of the potential barrier for $f=0$.
For the observables $\mathcal{C}_{S}$, $J_{S}$, $f$, and $\Delta\psi$, the unknown pair $\left(\beta, \Delta\overline{V}\right)$ can be inferred by using the following equation:
\begin{align}
\mathcal{C}_{S,i}
=\frac{2e^{\beta\left(\Delta\overline{V} + f_{i}\Delta\psi_{i}\right)}}{1+e^{\beta 2\pi f_{i}}}
~~(i=1,2).
\label{C_S,i}
\end{align}

We note that, in Eq.~(\ref{C_S,i}), each $f_{i}$ is obtained from the \textit{i}-th pair of observables ($J_{S}$, $P_{S}(\psi)$) as
\begin{align}
f_{i}
&= \langle d +H(\psi)\rangle_{\psi}
= \left<\left[J_{S}+\beta^{-1}P_{S}^{\prime}(\psi)\right]P_{S}(\psi)^{-1}\right>_{\psi} \notag \\
&= \left< J_{S}P_{S}(\psi)^{-1}\right>_{\psi},
\label{define_Q}
\end{align}
which is derived from the relationship
\begin{align}
d + H(\psi)
= -V^{\prime}(\psi)
= \left[J_{S}+\beta^{-1}P_{S}^{\prime}(\psi)\right]P_{S}(\psi)^{-1}
\label{-Vd}
\end{align}
for the stationary PDF $P_{S}(\psi)$ of Eq.~(\ref{eq:0428_partian_tPpsit}) and the identity $\left< P_{S}^{\prime}(\psi)P_{S}(\psi)^{-1}\right>_{\psi}=0$.

Now, we numerically verify \textit{Result 1} through the following example.

\textbf{Example.}
Following \cite{Pikovsky2015}, we consider the phase equation \eqref{eq:0209_dot_psi_5} with a {PSF}:
\begin{align}
Z(\theta)=2\sqrt{r}\cos{\theta}+2\sqrt{1-r}\cos{2\theta},
\label{Z_theta}
\end{align}
and an external forcing
\begin{align}
q(\theta)=a_{0}/2+a_{1}\cos{\theta}+b_{1}\sin{\theta}+a_{2}\cos{2\theta}+b_{2}\sin{2\theta},
\label{q_theta}
\end{align}
whose coefficients are randomly chosen such that $a_{0}/2 = -0.00152185$, $a_{1} = -0.12247362$, $b_{1} = -0.13656613$, $a_{2} = 0.03332079$, and $b_{2} = -0.21168956$.
This gives $\Vert q \Vert_2=0.5$. 
In this example, for $r=0.2$, $P_{S}(\psi)$ has two peaks (cf.~\cite{Speck2007}), and for $r=1.0$, $P_{S}(\psi)$ has one peak (cf. \cite{Blickle2007a, Blickle2007, Speck2006}).
In addition, for $r=0.2$, the upper limit of $f$, for which $V(\psi)$ has its extreme value, is $f_{\max} \simeq 0.2736$ and the lower limit is $f_{\min} \simeq -0.1979$.
For $r=1.0$, the upper limit is $f_{\max}\simeq 0.1834$ and the lower limit is $f_{\min}~(=-f_{\max})\simeq -0.1834$.

\textbf{Numerical verification for Result 1.}
For the above Example, from the SDE (\ref{eq:0428_dpsi/dtau}) with assumed values of $\left(\beta,\Delta\overline{V}\right)=(10.0,0.3152)$, two sets of the measured observables $\left(\widetilde{f_{i}}, \widetilde{\mathcal{C}_{S,i}}, \widetilde{\Delta\psi}_{i}\right)~(i=1,2)$ are obtained within reasonable accuracy for given (true) $f_{1}=0.28\times10^{-2}$ and $f_{2}=0.26\times10^{-1}$, respectively.
See Table~\ref{table_experiment2-1} in Appendix~\ref{sec:appendix_G} for more details.
Here we focus on the case of $f>0$, since the case of $f<0$ is treated similarly.

Solving Eq.~(\ref{C_S,i}) for $\left(\beta, \Delta\overline{V}\right)$ with the Newton--Raphson method, we found that the two sets of measured $\left(\widetilde{f_{i}}, \widetilde{\mathcal{C}_{S,i}}, \widetilde{\Delta\psi}_{i}\right)$ shown in Table~\ref{table_experiment2-1} of Appendix~\ref{sec:appendix_G} yield the estimated values $\left(\widetilde{\beta}, \widetilde{\Delta\overline{V}}\right) = (9.826, 0.3143)$ for $r=0.2$, which are close to the true values $(10.0,0.3152)$.
This estimation was verified to become more accurate if several sets of measurements are available e.g., by using least mean-square fitting.
For $r=1.0$, the same accuracy and tendency were verified (data not shown).

Despite these successful results, we need to be careful to avoid excessively large $\beta \vert f \vert$ when using the approximation (\ref{eq:0515_C_S});
there is a practical upper limit of $\beta \vert f \vert$ when using Eq.~(\ref{eq:0515_C_S}) for real applications.
A more detailed analysis of this approximation error is given in Appendix~\ref{sec:appendix_H}.

In the above Example, we inferred $\beta$ and $\Delta\overline{V}$ from two sets of observables using \textit{Result 1}.
We have another option for inferring $\beta$ and $\Delta V$ from only one set of observables; the analysis in Appendix~\ref{sec:appendix_I} provides the details of this option, although it shows an inferior result compared to the above one obtained by using \textit{Result 1} with two sets of observables.

\subsection{Inference of noise intensity from the measured potential}

Now, we consider the \textbf{Case (ii)}.
For a monotonic decreasing potential $V$ with $V^{\prime}(\psi)<0$ for $f \gg 1 > f_{\max}$, the following approximation holds from Eq.~(\ref{form_Ps}), up to the order of $\beta^{-3}$, for sufficiently large $\beta$:
\begin{align}
&P_{S}(\psi)
\approx C^{-1}\left\{-\beta^{-1}V^{\prime}(\psi)^{-1} - \beta^{-2}\left[V^{\prime\prime}(\psi)\left/ V^{\prime}(\psi)^{3}\right. \right]\right. \notag \\
&\left.-\beta^{-3}[3V^{\prime\prime}(\psi)^{2}/V^{\prime}(\psi)^{5}-V^{\prime\prime\prime}(\psi)/V^{\prime}(\psi)^{4}]\right\},
\label{eq:0523_P_S(PSI)}
\end{align}
which is obtained from \cite[Chap.~2, Sec.~1, Theorem~1, Eq.~(1.12)]{R.Wong1989}.
For the derivation of Eq.~(\ref{eq:0523_P_S(PSI)}), see Appendix~\ref{sec:appendix_E}.

Plugging $\psi = \psi_{\max,\min}$, respectively, into Eq.~(\ref{eq:0523_P_S(PSI)}) and using $V^{\prime}(\psi_{\max,\min}) = - J_{S}P_{S}(\psi_{\max,\min})^{-1}$, we have the following estimate of $\beta$, which we denote by $\beta_{A}$:
\begin{align}
\beta_{A}
\equiv\left(C_{\min,2}-C_{\max,2}\right)/\left(C_{\max,1}-C_{\min,1}\right),
\label{eq:0725_tild_beta}
\end{align}
where
\begin{align}
C_{\ast,1} &\equiv V^{\prime\prime}(\psi_{\ast})V^{\prime}(\psi_{\ast})^{-2},
~\text{and} \notag \\
C_{\ast,2} &\equiv 3V^{\prime\prime}(\psi_{\ast})^{2}V^{\prime}(\psi_{\ast})^{-4}-V^{\prime\prime\prime}(\psi_{\ast})V^{\prime}(\psi_{\ast})^{-3}
\label{C_ast_1,2}
\end{align}
with the suffix $\ast$ representing $\max$ or $\min$.
For the derivation of Eqs.~(\ref{eq:0523_P_S(PSI)}) and (\ref{eq:0725_tild_beta}), see Appendix~\ref{sec:appendix_E}.

As mentioned in Sec.~\ref{sec:2}, $V(\psi)$ is indirectly obtained from $\beta$, $J_S$, and $P_S(\psi)$ by the method in \cite{Udo2012,Blickle2007a,Blickle2007,Speck2006,Speck2007}
for microsystems of colloidal particles. 
In contrast, for noisy limit-cycle oscillators, this $V(\psi)$ is obtained from $H(\psi)$ measured by the method in \cite{Miyazaki2006,Stankovski2017}.
Therefore, we can obtain $\beta$ by Eqs.~(\ref{eq:0725_tild_beta}) and (\ref{C_ast_1,2}) for the measured $V(\psi)$ and $P_S(\psi)$.
This is the second main result in this study and can be summarized as follows:

\textbf{\textit{Result 2}.}
Suppose that $V(\psi_{\max,\min})$, $V^{\prime}(\psi_{\max,\min})$, $V^{\prime\prime}(\psi_{\max,\min})$, and $V^{\prime\prime\prime}(\psi_{\max,\min})$ are obtained from measured $\widetilde{V}(\psi)$, for instance, by the method in \cite{Miyazaki2006,Stankovski2017}, where $\psi_{\max}$ and $\psi_{\min}$ are measured from the observable $P_{S}(\psi)$.
In this setting, $\beta$ can be inferred by ${\beta_A}$ of Eq.~(\ref{eq:0725_tild_beta}) from the measured observables $V(\psi)$ and $P_{S}(\psi)$.

\textbf{Numerical verification for Result 2.}
First, we verify the accuracy of $\beta_A$ obtained by using \textit{Result 2} from the observables $V(\psi)$ and $P_S(\psi)$.
For the present example, Eq.~\eqref{Z_theta} with $r=0.2$ and $f=2.8$, the estimation errors of $\beta_{A}~(=(\beta - \beta_{A})/\beta)$ are quite low, i.e., $<0.1\%$ for sufficiently large $\beta$ (see Table~\ref{tab:estimate_beta=2} in Appendix~\ref{sec:appendix_G}).

Next, we consider the sensitivity of $\beta_A$ to the measurement error.
The measured $V(\psi)$ contains a certain amount of error with respect to the true $V(\psi)$ since, in our setting, $V$ is obtained as $V(\psi) = - \int^{\psi}d\psi^{\prime}\left[d+H(\psi^{\prime})\right]$, where $d$ is supposed to be known and $H(\psi^{\prime})$ can be experimentally determined (cf.~\cite{Miyazaki2006,Stankovski2017}).
For details of this experimentally determined coupling function $H(\psi)$, see Appendix~\ref{sec:appendix_J}.
Thus, with respect to a certain amount of variation (measurement error) in $V$, it is practically necessary to consider the sensitivity of the accuracy of the estimator $\beta_A$.

\begin{figure}[h]
\begin{center}
\includegraphics[scale=0.25]{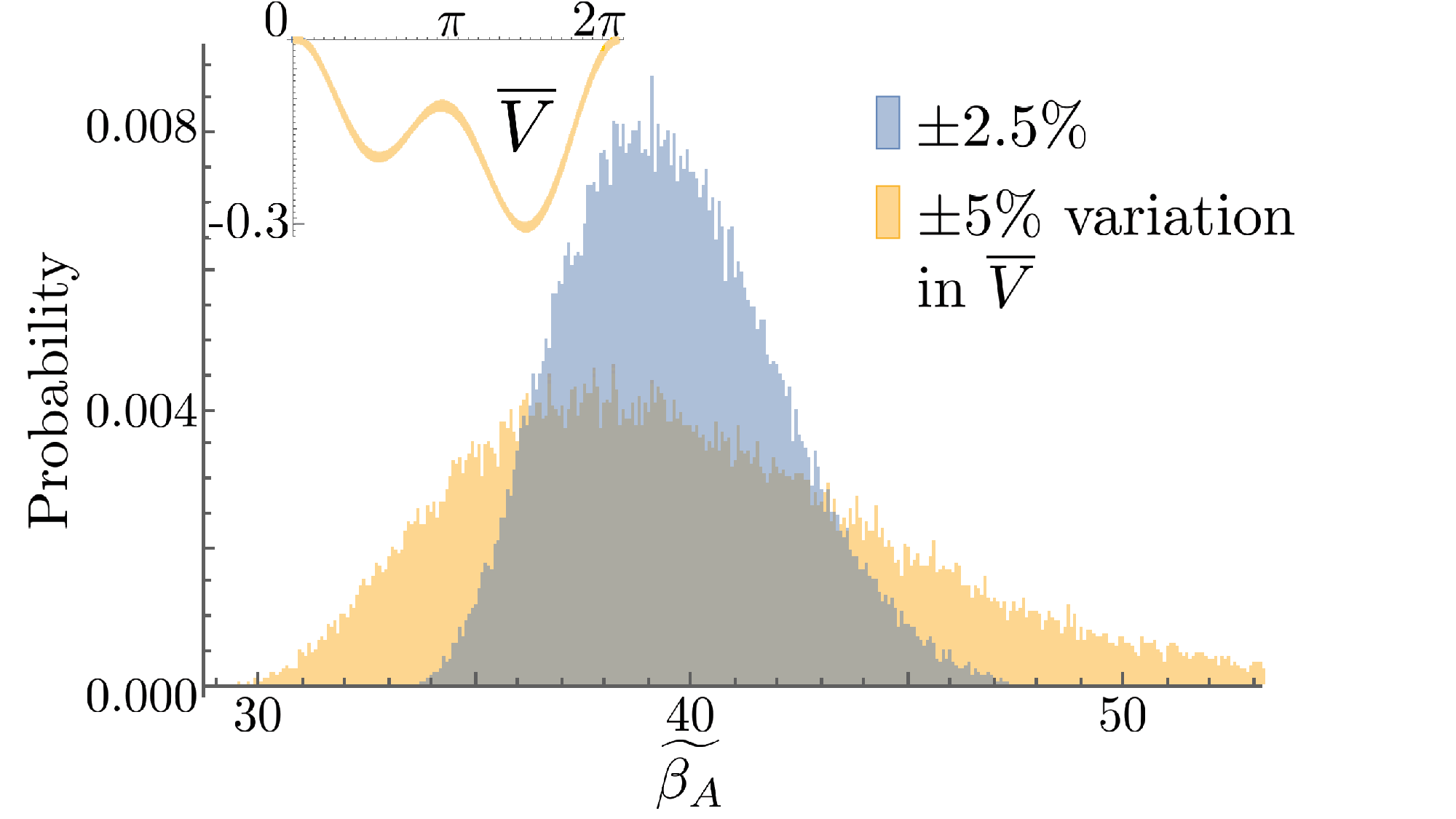}
\end{center}
\vspace{-0.5cm}
\caption{Distributions of the estimate $\widetilde{\beta_{A}}$ around true $\beta=40.0$.
For $r=0.2$ and $f=1.0$ in the Example, $10^{5}$ variations of $V$ were generated uniformly randomly within $\pm2.5\%$ and $\pm5\%$ error with respect to the original $V$.
The inset shows the $\pm5\%$ variations of the purely periodic part $\overline{V}(\psi)$.
}\label{fig0710_hist}
\end{figure}

Figure~\ref{fig0710_hist} shows that the distribution of the estimate $\widetilde{\beta_A}$ peaks around the true $\beta~(=40.0)$.
The shape of the distribution becomes asymmetric for a greater variation in $V$.
The width of the distribution is somewhat broad even for a $\pm 2.5\%$ variation in $V$ with all its Fourier coefficients being randomly perturbed (uniformly within $\pm 2.5\%$).
This broadness (i.e., sensitivity) is due to the particular structure of Eqs.~(\ref{eq:0725_tild_beta}) and (\ref{C_ast_1,2}); 
as listed in Table~\ref{tab:estimate_beta=2} in Appendix~\ref{sec:appendix_G}, $C_{\ast,1}$ in the denominator of Eq.~(\ref{eq:0725_tild_beta}) is on the order of $10^{-4}$--$10^{-3}$ with respect to the order of $10^{-2}$ in the numerator $C_{\ast,2}$.
This denominator can be sensitive to a small variation (measurement error) of the value of $V^{\prime\prime}(\psi_{\ast})~ \left(\text{approximately on the order of}~10^{-1}\right)$ at $\psi_{\ast}=\psi_{\max,\min}$, in $C_{\ast,1}$.
It is not straightforward to explain the asymmetric shape of the distribution since Eqs.~(\ref{eq:0725_tild_beta}) and (\ref{C_ast_1,2}) depend on $V^{\prime}(\psi_{\ast})$, $V^{\prime\prime}(\psi_{\ast})$, and even $V^{\prime\prime\prime}(\psi_{\ast})$ in a complex manner, each of which affects the estimation error.

In \textit{Result 2}, it is reasonable to ask what happens if only $P_{S}$ is measured but $V$ is not available.
The answer to this question is somewhat negative; the analysis in Appendix~\ref{sec:appendix_F} provides the reason to some extent.

\subsection{Inference of phase response from the estimated noise intensity}

Having obtained the estimate $\beta$ for $\vert f \vert \ll f_{\max}$ (or $\vert f_{\min} \vert$) and $\vert f \vert \gg f_{\max}$ (or $\vert f_{\min} \vert$) in \textit{Result 1} and \textit{Result 2}, respectively, the phase response, i.e., the PSF $Z(\theta)$ can be inferred as follows.
Here, we represent $q(\theta)$ and $Z(\theta)$ in Eq.~(\ref{define_Gamma}) in Fourier series with their coefficients ($a_n$, $b_n$) and ($c_n$, $d_n$):
$q(\theta)=a_{0}/2+\displaystyle\sum_{n\ge1}\left(a_{n}\cos{n\theta}+b_{n}\sin{n\theta}\right)$ and $Z(\theta)=c_{0}/2+\displaystyle\sum_{n\ge1}\left(c_{n}\cos{n\theta}+d_{n}\sin{n\theta}\right)$, respectively.

First, for $n=0$, from Eqs.~(\ref{define_Gamma}) and (\ref{define_Q}), 
\begin{align}
f=d + Z_0q_0
\label{slope}
\end{align}
is obtained, where $Z_0~(=c_{0}/2)$ and $q_0~(=a_{0}/2)$ are the constant part of $Z(\theta)$ and $q(\theta)$, respectively.
Thus, $Z_0$ is determined by measuring $f$, $d$, and $q_0$.

Next, for $n \geq 1$, ($c_n$, $d_n$) are determined as follows.
We represent $H(\psi)$ in Eq.~(\ref{define_Gamma}) in Fourier series:
$H(\psi)=g_{0}/2+\displaystyle\sum_{n\ge1}\left(g_{n}\cos{n\psi}+h_{n}\sin{n\psi}\right)$, and the Fourier coefficients ($g_n$, $h_n$) are obtained from the measured $H(\psi)$ through Eq.~(\ref{-Vd}) or directly measured by using the method in \cite{Stankovski2017, Miyazaki2006}.
These ($g_n$, $h_n$) are related to $q$ and $Z$ with Eq.~(\ref{define_Gamma}) as
\begin{align}
g_{n} = (a_{n}c_{n} + b_{n}d_{n})/2,~
h_{n} = (a_{n}d_{n} - b_{n}c_{n})/2.
\label{eq:0523_alpha_nbeta_n}
\end{align}
Thus, if the Fourier coefficients ($a_n$, $b_n$) of the external forcing $q$ are measured or already known, as in the Example, then ($c_n$, $d_n$) of $Z$ are determined from Eq.~(\ref{eq:0523_alpha_nbeta_n}).
This is the third main result in this study and can be summarized as follows:

\textbf{\textit{Result 3}.}
For $\vert f\vert\ll f_{\max}$ (or $\vert f_{\min}\vert$), suppose that we have the measured $J_S$, $P_{S}(\psi)$, and the inferred $\widetilde{\beta}$ from \textit{Result 1}, and the information of $d$ and $q(\Omega t)$.
Then, $Z(\theta)$ is determined from Eqs.~(\ref{slope}) and (\ref{eq:0523_alpha_nbeta_n}).
In contrast, for $\vert f\vert\gg f_{\max}$ (or $\vert f_{\min}\vert$), $Z(\theta)$ is directly determined from Eqs.~(\ref{slope}) and (\ref{eq:0523_alpha_nbeta_n}) using the measured $H(\psi)$ with the method in \cite{Stankovski2017, Miyazaki2006}, where the inferred $\widetilde{\beta}$ from \textit{Result 2} is not involved.

\textbf{Numerical verification for Result 3.}
First, for the Example of Eq. \eqref{Z_theta} with $r=0.2$, $\beta=10.0$, and $f=0.28\times10^{-1}$, we measured $J_{S}$ and $P_{S}(\psi)$ from the SDE (\ref{eq:0428_dpsi/dtau}) and obtained $\widetilde{\beta}~(\simeq 9.826)$, as presented in Numerical verification for Result 1.

Next, for these observables, $J_{S}$ and $P_{S}(\psi)$, and from the inferred $\widetilde{\beta}$, we measured $d+H(\psi)$ from Eq.~(\ref{-Vd}) and obtained its Fourier series.
Then, we measured ${\widetilde{f}}$ from Eq.~(\ref{define_Q}) and compared it to the true value of $f$, which gives $\widetilde{f}\simeq0.277\times10^{-1}$ with an error of $1.071\%$ from the true $f=0.28\times10^{-1}$.

\begin{figure}[htb]
\includegraphics[scale=0.4]{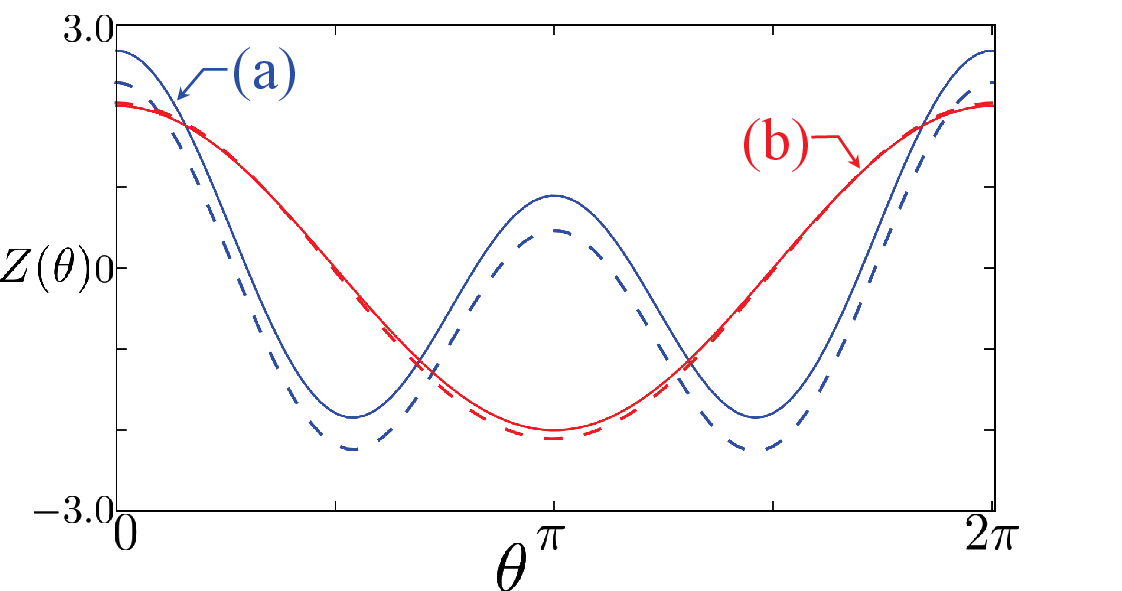}
\label{fig0523_1c}
\caption{Comparison between the true $Z(\theta)$ (solid line) and the inferred one (dashed line)
for (a) $r=0.2$, $\widetilde{f}\simeq0.277\times10^{-1}$, and $\widetilde{\beta}=9.826$ 
and (b) $r=1.0$, $\widetilde{f}\simeq0.195\times10^{-2}$, and $\widetilde{\beta}=9.619$.
See Appendix~\ref{sec:appendix_G} for details. 
}\label{fig0523_1}
\end{figure}

Finally, from Eq.~(\ref{slope}), the estimate of $Z_0$ is obtained for a given $d~\left(=0.28\times10^{-1}\right)$ and $q_0~\left(=-0.152185\times10^{-2}\right)$, and the Fourier coefficients ($c_n$, $d_n$) of $Z$ are obtained for $n=1,2$ from Eq.~(\ref{eq:0523_alpha_nbeta_n}), as shown in Table~\ref{table_0612_Fourier_coefficients} in Appendix~\ref{sec:appendix_G}.
Thus, $Z$ is reconstructed as shown in Fig.~\ref{fig0523_1} within a certain accuracy.
In addition to the above case of $r=0.2$, a similar result is obtained for the case of $r=1.0$ with different $\beta$ and $f$, which is also shown in Fig.~\ref{fig0523_1}.

Regarding \textit{Result 3}, the requirement for inferring $Z_0$ correctly is that $\vert q_0\vert$ should not be 0 or too small.
Otherwise, the accuracy of the inferred $Z_0$ degrades as $\vert q_0 \vert$ gets smaller since $Z_0$ becomes sensitive to the measurement error of $f$ and $d$ from Eq.~(\ref{slope}).

\section{\label{sec:discussions}Discussion and open problems}

In this study, we developed a framework for inferring the effective noise intensity and PSF of noisy synchronous oscillators using their stationary PDFs, probability currents, and coherence measures.
Through asymptotic approximations, we derived three useful results that relate experimentally measurable quantities to the key informations of the reduced phase equation of the system;
the effective noise intensity and PSF.
Our results are numerically verified using a simple example of noisy synchronous oscillators.
Regarding \textit{Result 1} and \textit{Result 2}, to the best of the authors' knowledge, the simple asymptotic formulas (\ref{C_S,i}) or Eqs.~(\ref{eq:0725_tild_beta}) and (\ref{C_ast_1,2}) have not yet been documented despite the long history of the Fokker--Planck equation and Brownian motors (cf.~\cite{Risken1996} and \cite{Reimann2002}, respectively, for instance).

In addition to the presented analytical framework, if all the observables are available with relatively high accuracy, Eqs.~(\ref{form_Ps}--\ref{form_C}) can be numerically solved for the noise intensity in a different way;
this numerical option can be applied for any magnitude of the noise intensity and the potential slope.
However, this method seems too sensitive for measurement errors in the observables to correctly infer the noise intensity.
We intend to report relevant work elsewhere.

Recent theoretical studies have made it possible to consider forcing with a larger magnitude by introducing the phase-amplitude coordinate (i.e., augmented phase reduction \cite{Wilson2022,Monga2019} and the phase-amplitude reduction \cite{Shirasaka2017}) in conjunction with the phase $\phi$.
Although this is beyond the scope of the present study using a phase-only description of the oscillator, an extension of our framework for these coordinates is left as a future study.

We now discuss three possible applications of the framework presented in this study.
We consider examples of the rotating Stirling heat engine \cite{Kitaya2022,Izumida2016,Toyabe2020,Tokuda2017,Tokuda2014,Izumida2023}, the electrochemical oscillator \cite{Zachary2020}, and semiclassical quantum limit-cycle oscillators \cite{Knunz2010, Kato2019, Sadeghpour2013, Noh2020, Lifshitz2021}.
Our framework expects the following intriguing future applications for these systems.

(i)
Application to rotating Stirling heat engines.
Recently, a series of interesting results have been reported for the Stirling engine (SE), in which the SE has been shown to be a limit-cycle oscillator \cite{Izumida2016,Toyabe2020}.
In addition to an experimental study on a macroscale SE \cite{Toyabe2020}, a mesoscale SE \cite{Kitaya2022} was numerically investigated, and mutual synchronization of two coupled SEs has been experimentally \cite{Tokuda2014} and theoretically \cite{Izumida2023} investigated.
Our framework may be used to infer the noise intensity and PSF for these SEs.
An important question is how the nonlinear dynamics of the SE and the environmental (thermal) noise determine the noise intensity in our framework\RED{,} since noise intensity is related to the energetics of micro/mesoscale systems (including the SE) in stochastic thermodynamics \cite{Udo2012,Ciliberto2017}.

(ii)
Application to electrochemical oscillators.
Ref. \cite{Zachary2020} theoretically showed that two uncorrelated noise sources independently injected into two mutually coupled oscillators can be more effective than a common noise source in enhancing synchronization if the noise intensity is neither too weak nor too strong.
Notably, for a certain intermediate intensity of white Gaussian noise, the electrochemical experiments in \cite{Zachary2020} have demonstrated this phenomenon.
Our results can be used to infer the specific intermediate intensity of noise for enhancing mutual synchronization.
Since the electrochemical system is a typical experimental system that can exhibit noisy limit-cycle oscillations, which also underlie the functioning of noisy biological oscillations, our framework has the potential to clarify the interplay of noise and coupling for a variety of systems.

(iii)
Application to semiclassical quantum limit-cycle oscillators.
Optical and nanomechanical quantum limit-cycle oscillators \cite{Knunz2010} have been discussed using a quantum master equation \cite{Kato2019, Sadeghpour2013, Noh2020, Lifshitz2021}, which can be approximated by a classical Fokker--Planck equation under the semiclassical approximation.
Moreover, by phase reduction~\cite{Kato2019}, the system can be described by the SDE (\ref{eq:0428_dpsi/dtau}) for the oscillator phase.
Thus, the framework presented for inferring noise intensity and PSF can also be applicable to analyze quantum limit-cycle oscillators in the semiclassical regime.
In particular, when the accurate model of the experimental system is unavailable, our framework may be used to infer the effective intensity of the quantum noise and PSF of the system.
Thus, it may be used to analyze noisy synchronization in general semiclassical quantum oscillators.

The above applications of our framework will be considered in future studies.

\begin{acknowledgments}

Two of the authors (H.~T. and S.~S.) would like to thank Yoshihiko Hasegawa for his beneficial comments and critical reading of the manuscript.
They would also like to thank Shota Izumi and Chikara Furukawa for their help during preparation of the manuscript.
One of the authors (H.~T.) acknowledges the useful discussions he had with Arkady Pikovsky, Toru Ohira, Naoko Nakagawa, Isao Tokuda, Yoshihiko Hasegawa, and Yuzuru Kato.
H.~T. would like to dedicate this work to the memory of A.~J. Lichtenberg and the discussions they had on the Fokker--Planck coefficients \cite{Lieberman1972} and inverse temperature.
This work was supported in part by a Japan Society for the Promotion of Science (JSPS) KAKENHI Grant-in-Aid for Scientific Research (B) (grant number JP24K03008).
H.~N. acknowledges JSPS KAKENHI (JP22K11919, JP22H00516) and the Japan Science and Technology Agency CREST (JP-MJCR1913) for their financial support.
Y.~J. acknowledges JSPS KAKENHI (JP23H01409) for its financial support.

\end{acknowledgments}

\appendix
\section{\label{sec:appendix_A}Derivation of Eq.~(\ref{eq:0515_C_S})}

For the lowest-order asymptotic expansion of $\mathcal{C}_{S}~\left(=P_{S}(\psi_{\max})/P_{S}(\psi_{\min})\right)$ in Eq.~(\ref{eq:0515_C_S}) in the main text, we apply Laplace's approximation (Laplace's formula, cf.~\cite{R.Wong1989}, p. 55) to $P_S(\psi_{\min})$ and $P_S(\psi_{\max})$, respectively, as follows.
For simplicity, we introduce, as depicted in Fig.~\ref{fig_potential}, the coordinate $\psi$ with its origin $\psi=\psi_{u}$ ($\rightarrow \psi_{\min}$ as $\vert f \vert \rightarrow0$), and then from Eq. (\ref{vpsi_fpsi}) $V(\psi_{u})=\overline{V}(\psi_{u})-f\cdot0=V_{\min}$.

First, for $P_{S}(\psi_{\min})$, from Eq.~(\ref{form_Ps}) and $V(\psi)=\overline{V}(\psi)-f\psi$, it follows that
\begin{align}
&P_{S}(\psi_{\min})
=C^{-1}e^{-\beta {V}(\psi_{\min})} \int_{\psi_{\min}}^{\psi_{\min}+2\pi}  e^{\beta V(\psi^{\prime})} d\psi^{\prime} \notag \\
&~~=C^{-1}e^{-\beta V(\psi_{\min})}\int_{\psi_{\min}}^{\psi_{\min}+2\pi}e^{-\beta  f \psi^{\prime}}e^{\beta  \overline{V}(\psi^{\prime})} d \psi^{\prime}.
\label{eq:0118_P_Spsi_min}
\end{align} 
Here, we assume, without loss of generality, for any positive $f~(<f_{\max}\ll1)$, $V(\psi^{\prime})$ has a maximum at $\psi^{\prime}=\psi_{u}$. 
Then, $\psi_u \approx \psi_{\min}$ (i.e., $\psi_u \rightarrow \psi_{\min}$ as $f\rightarrow0$) since $f \ll 1$ and $\beta \gg 1$, as shown in Eq.~(\ref{eq:psi_max-min}).
This implies $\overline{V}(\psi^{\prime})$ becomes maximal at both ends of $\left[\psi_{\min},\psi_{\min}+2\pi \right]$, as $f\rightarrow0$ (which yields `$\times (1/2)$' in Eq.~(\ref{eq:0626_integral(4)}) below).
Here, we note that $\overline{V}^{\prime\prime}(\psi_{\min}+2\pi)=\overline{V}^{\prime\prime}(\psi_{\min})$.
Then, by Laplace's formula, 
\begin{align}
&\int_{\psi_{\min}}^{\psi_{\min}+2\pi} e^{-\beta f \psi^{\prime}}e^{\beta \overline{V}(\psi^{\prime})}d\psi^{\prime}\notag \\
&\approx e^{-\beta f \psi_{\min}}\sqrt{\frac{2\pi}{-\beta \overline{V}^{\prime\prime}(\psi_{\min})}}
e^{\beta \overline{V}( \psi_{\min})}
\times (1/2) \notag \\ 
&+ e^{-\beta f \left(\psi_{\min}+2\pi\right)}
\sqrt{\frac{2\pi}{-\beta \overline{V}^{\prime\prime}(\psi_{\min})}}
e^{\beta \overline{V}( \psi_{\min})} \times (1/2) \notag \\
&=\frac{1+e^{-\beta 2\pi f}}{2}
\sqrt{\frac{2\pi}{-\beta V^{\prime\prime}(\psi_{\min})}}
e^{\beta V(\psi_{\min})}, \notag \\
& \hspace{4.5cm} \beta\rightarrow\infty,~f\rightarrow0.
\label{eq:0626_integral(4)}
\end{align} 
Then, substituting Eq.~(\ref{eq:0626_integral(4)}) in Eq.~(\ref{eq:0118_P_Spsi_min}) gives:
\begin{align}
P_{S}(\psi_{\min})
\approx C^{-1}\frac{1+e^{-\beta 2\pi f}}{2}\sqrt{\frac{2\pi}{-\beta V^{\prime\prime}(\psi_{\min})}},& \notag \\
\beta\rightarrow\infty,~f\rightarrow0.&
\label{eq:P_S_psi_min}
\end{align}

Next, for $P_S(\psi_{\max})$, an asymptotic approximation similar to Eq.~(\ref{eq:P_S_psi_min}), is obtained in the following manner.
Since $\psi_{s,u}\rightarrow\psi_{\max,\min}~(f\rightarrow0)$, respectively, as shown in Fig. \ref{fig_potential} we define $\Delta\psi\equiv \psi_{s}-\psi_{u}~(\rightarrow \psi_{\max}-\psi_{\min})$ and $\Delta V\equiv V_{\min}-V_{\max}+f\Delta\psi$, which gives
\begin{subequations}
\begin{align}
&V(\psi_{\max})
= \overline{V}(\psi_{\max}) - f\cdot\left(\psi_{\max} - \psi_{\min}\right) \notag \\
&\hspace{1.3cm}\rightarrow V_{\max} - f\Delta\psi,
\label{} \\
&V(\psi_{\min} + 2\pi)
\rightarrow V_{\min} - 2\pi f
\hspace{0.5cm}(f\rightarrow0).
\label{}
\end{align}
\label{Varrow}
\end{subequations}

Since $\psi_{\min}+2\pi \in \left[\psi_{\max},\psi_{\max}+2\pi\right]$, by Laplace's formula,
\begin{align}
\int_{\psi_{\max}}^{\psi_{\max}+2\pi}e^{\beta V(\psi^{\prime})}d\psi^{\prime}
\approx \sqrt{\frac{2\pi}{-\beta V^{\prime\prime}(\psi_{\min})}}
e^{\beta V(\psi_{\min}+2\pi)}.
\label{eq:1208_P_S_psi_max}
\end{align}
Substituting this and Eq.~(\ref{Varrow}) in Eq.~(\ref{form_Ps}) and $V_{\min}-V_{\max} + f \Delta \psi = \Delta V$ gives
\begin{align}
&P_{S}(\psi_{\max})\approx C^{-1} e^{-\beta V(\psi_{\max})}\sqrt{\frac{2\pi}{-\beta V^{\prime\prime}(\psi_{\min})}}e^{\beta V(\psi_{\min}+2\pi)} \notag \\
&~~\approx C^{-1}\sqrt{\frac{2\pi}{-\beta V^{\prime\prime}(\psi_{\min})}}
e^{-\beta (V_{\max}- f \Delta\psi)} \times e^{\beta (V_{\min}- 2\pi  f)} \notag \\
&~~= C^{-1}\sqrt{\frac{2\pi}{-\beta V^{\prime\prime}(\psi_{\min})}}
e^{\beta(\Delta V-2\pi f)}.
\label{eq:1208_P_S_psi_min}
\end{align}

Thus, from Eq.~(\ref{eq:1208_P_S_psi_max}) and Eq.~(\ref{eq:1208_P_S_psi_min}), 
\begin{align}
\mathcal{C}_{S}&=P_{S}(\psi_{\max})/P_{S}(\psi_{\min}) 
\approx \frac{2}{1+e^{-\beta 2\pi f}}e^{\beta(\Delta V-2\pi f)} \notag \\
&=\frac{2}{1+e^{\beta 2\pi f}}e^{\beta\Delta V},
\hspace{1.5cm}\beta\rightarrow\infty,~f\rightarrow0.
\label{eq:1128_C_S}
\end{align}

\setcounter{equation}{1}

\section{\label{sec:appendix_B}Derivation of Eqs.~(\ref{eq:C_D_0729}) and (\ref{eq:0515_C_D})}

For the Laplace-type integral $F(\beta) \equiv \int\int_{\Omega}e^{\beta h(x,y)}g(x,y)dxdy$, we assume that
(i) an interior, nondegenerate maximum point ($x_{0}$, $y_{0}$) exists in the convex region $\Omega$, and
(ii) at ($x_0$, $y_0$), $\nabla h(x_0,y_0)~\left(=\left(\frac{\partial h}{\partial x}, \frac{\partial h}{\partial y} \right)|_{(x_0,y_0)}\right) =(0,0)$, and the Hessian $\vert Hh(x_0,y_0)\vert$ is positive definite.
Then, the following holds (cf. \cite[Chap. 8, Sec. 10, Eq. (10.10)]{R.Wong1989}):
\begin{align}
F(\beta) \approx \frac{2\pi g(x_0,y_0)}{\beta \sqrt{\vert Hh(x_0,y_0) \vert}}e^{\beta h(x_0,y_0)},
\hspace{1cm}\beta\rightarrow\infty.
\label{eq:0207_f(beta)}
\end{align}

Then, by the formula (\ref{eq:0207_f(beta)}), for $\beta\rightarrow\infty$ and $f\rightarrow0$, Eq.~(\ref{D_D_0}) is approximated in the following manner.
First, the numerator of Eq.~(\ref{D_D_0}) is
\begin{widetext}
\begin{align}
&D_0\langle I_{+}(\psi)^{2}I_{-}(\psi)\rangle_{\psi}
=D_0\int_{\psi_{0}}^{\psi_{0}+2\pi}\int_{0}^{2\pi}\int_{0}^{2\pi}\int_{0}^{2\pi}e^{\beta h(\psi,\psi^{\prime},\psi^{\prime\prime},\psi^{\prime\prime\prime})}d \psi^{\prime\prime\prime} d\psi^{\prime\prime} d \psi^{\prime} d\psi \notag \\
&=D_0\int_{\psi_{0}}^{\psi_{0}+2\pi}d\psi \int_{\psi}^{\psi-2\pi} d\psi_{1} \int_{\psi}^{\psi-2\pi} d\psi_{2} \int_{\psi}^{\psi+2\pi} d\psi_{3} 
e^{\beta \overline{h}(\psi,\psi_{1},\psi_{2},\psi_{3})}
\frac{\partial (\psi,\psi^{\prime},\psi^{\prime\prime},\psi^{\prime\prime\prime})}{\partial(\psi,\psi_{1},\psi_{2},\psi_{3})}\notag \\
&=D_0\iint_{\Omega_{1}}d\psi d\psi_{1} e^{\beta h_{1}(\psi,\psi_{1})}
\times \iint_{\Omega_{2}}d\psi_{2}d\psi_{3} e^{\beta h_{2}(\psi_{2},\psi_{3})}\times 1 \notag \\
&\approx D_0\left\{\frac{2\pi}{\beta}\left[ -V^{\prime\prime}(\psi_{\max})V^{\prime\prime}(\psi_{\min}) \right]^{-\frac{1}{2}}e^{\beta \Delta V}\right\}^{2},
\label{eq:0623_I+^2I_}
\end{align}
\end{widetext}
where
$\Omega_{1}\equiv[\psi_{0},\psi_{0}+2\pi]\times[\psi,\psi-2\pi]$,
$\Omega_{2}\equiv[\psi,\psi-2\pi]\times[\psi,\psi+2\pi]$,
$\overline{h}(\psi,\psi_{1},\psi_{2},\psi_{3})\equiv h_{1}(\psi,\psi_{1})h_{2}(\psi_{2},\psi_{3})$,
$h_{1}(\psi,\psi_{1})\equiv\overline{V}(\psi) - \overline{V}(\psi_{1}) + f\RED{\cdot}(\psi_{1} - \psi)$,
$h_{2}(\psi_{2},\psi_{3})\equiv\overline{V}(\psi_{3})-\overline{V}(\psi_{2}) + f\RED{\cdot}(\psi_{2}-\psi_{3})$
with $\psi_{1} \equiv \psi - \psi^{\prime}$,
$\psi_{2} \equiv \psi - \psi^{\prime\prime}$,
and $\psi_{3} \equiv \psi + \psi^{\prime\prime\prime}$.
In Eq.~(\ref{eq:0623_I+^2I_}), the following equation is obtained:
\begin{align}
\frac{\partial (\psi,\psi^{\prime},\psi^{\prime\prime},\psi^{\prime\prime\prime})}{\partial(\psi,\psi_{1},\psi_{2},\psi_{3})}=
\begin{vmatrix}
1	&	0	&	0	&	0	\\
1	&	-1	&	0	&	0	\\
1	&	0	&	-1	&	0	\\
-1	&	0	&	0	&	1	\\
\end{vmatrix}
=1,
\end{align}
which corresponds to ``$\times 1$'' in the third line of Eq.~(\ref{eq:0623_I+^2I_}).
In addition, in Eq.~(\ref{eq:0623_I+^2I_}), $h_{1}(\psi,\psi_{1})$ has the maximum value $\Delta V$ at $(\psi,\psi_{1})=(\psi_{s},\psi_{u})\rightarrow(\psi_{\max},\psi_{\min})\in\Omega_{1}$ and $h_{2}(\psi_{2},\psi_{3})$ also has the maximum value $\Delta V$ at $(\psi_{2},\psi_{3})=(\psi_{s},\psi_{u})\rightarrow(\psi_{\max},\psi_{\min})\in\Omega_{2}$ as $f \rightarrow 0$.

Next, the denominator of Eq.~(\ref{D_D_0}) is
\begin{align}
&\langle I_{+}(\psi)\rangle_{\psi}^3
~\left(=C^3\right) \notag \\
&\approx \left\{ \frac{2\pi}{\beta} \left[-V^{\prime\prime}(\psi_{\max})V^{\prime\prime}(\psi_{\min}) 
\right]^{-\frac{1}{2}}e^{\beta \Delta V} \right\}^{3},
\label{cubic_intI+}
\end{align}
which is obtained in Appendix~\ref{sec:appendix_D} below.

Then, substituting Eq.~(\ref{eq:0623_I+^2I_}) and Eq.~(\ref{cubic_intI+}) in Eq.~(\ref{D_D_0}) yields
\begin{align}
\mathcal{C}_{D}&=D_{0}/D
=\langle I_{+}(\psi)\rangle^{3}_{\psi} / \langle I_{+}^{2}(\psi)I_{-}(\psi)\rangle_{\psi} \notag \\
&\approx \frac{2\pi}{\beta} \left[-H^{\prime}(\psi_{\max})H^{\prime}(\psi_{\min}) \right]^{-\frac{1}{2}}e^{\beta\Delta V}, \notag \\
&\hspace{3.4cm}\beta\rightarrow\infty,~f\rightarrow0,
\label{approxC_D}
\end{align}
where $H^{\prime}(\psi) = -V^{\prime\prime}(\psi)$.

\section{\label{sec:appendix_C_0716}Relationship between Eq.~(\ref{eq:0515_C_S}) and Eq.~(\ref{eq:0515_C_D}) in the weak noise limit}
From Eq.~(\ref{eq:0515_C_D}) with a positive constant $\overline{\mathcal{C}}_{D}\equiv 2\pi \left[-H^{\prime}(\psi_{\max})H^{\prime}(\psi_{\min})\right]^{-\frac{1}{2}}$, for a sufficiently large $\beta$, we have
\begin{align}
\beta^{-1}\ln\mathcal{C}_D
&\approx\beta^{-1}\ln{\overline{\mathcal{C}}_{D}} + \beta^{-1}\ln\beta^{-1} + \Delta V \notag \\
&\rightarrow \Delta V~~(\beta\rightarrow\infty).
\label{}
\end{align}
On the other hand, from Eq.~\eqref{eq:0515_C_S} and $\beta^{-1}\ln\left[2/\left(1+e^{\beta 2\pi\vert f \vert}\right)\right] \rightarrow -2\cdot2\pi\vert f \vert~(\beta \rightarrow\infty)$, we obtain
\begin{align}
\beta^{-1}\ln{\mathcal{C}_{S}}
&\approx \beta^{-1}\left\{\beta\Delta V + \ln{\left[2/\left(1+e^{\beta2\pi\vert f \vert}\right)\right]}\right\} \notag \\
&\rightarrow \Delta V - 2\cdot2\pi\vert f \vert~~(\beta \rightarrow \infty). 
\label{}
\end{align}
This implies that the difference between $\beta^{-1}\ln \mathcal{C}_D$ and $\beta^{-1}\ln \mathcal{C}_S$ tends toward the constant $-4\pi\vert f \vert$, where the contribution of $H^{\prime}$ in Eq.~(\ref{eq:0515_C_D}) disappears in the weak noise limit.

\section{\label{sec:appendix_D}Asymptotic approximation of Eq.~(\ref{form_C})}

The asymptotic approximation of $C~(=\langle I_{\pm}\rangle_{\psi})$ is obtained in the following manner.
First, from Eq.~(\ref{form_C})
\begin{align}
C
&= C\int_0^{2\pi} P_{S}(\psi)d\psi \notag \\
&= \int_0^{2\pi}\int_{\psi}^{\psi +2\pi} e^{-\beta V(\psi)}e^{\beta V(\psi^{\prime})}d\psi^{\prime} d\psi.
\end{align}

Next, we use Eq.~(\ref{eq:0207_f(beta)});
for the Laplace-type integral in Appendix \ref{sec:appendix_B}, the range of integration $\mathrm{\Omega}\equiv\{(\psi,\psi^{\prime})|\psi \in [0,2\pi], \psi^{\prime} \in [\psi,\psi+2\pi]\}$ is a convex region (parallelogram), $h(x,y)=V(\psi)-V(\psi^{\prime})$, and $g(x,y)=1$.
In addition, $(x_0,y_0)~\left(=(0,0)\right)$ amounts to $(\psi,\psi^{\prime})~\left(=(\psi_{s},\psi_{u})\right)$, which satisfies the condition (i) in Appendix~\ref{sec:appendix_B}.
Furthermore,
$\vert Hh(x_0,y_0)\vert=
\begin{vmatrix}
\overline{V}^{\prime\prime}(\psi_{s})	&	0	\\
0	&	-\overline{V}^{\prime\prime}(\psi_{u})	\\
\end{vmatrix}$
is positive definite, and the condition (ii) is also satisfied.
Therefore, from Eq.~(\ref{eq:0207_f(beta)}), with $g(x_0,y_0)=1$ and $h(x_0,y_0)=V(\psi_s)-V(\psi_u)\rightarrow V(\psi_{\min})-V(\psi_{\max})=-\Delta V$, and noting that $\overline{V}^{\prime\prime}=V^{\prime\prime}$, the following asymptotic approximation is obtained:
\begin{align}
C\approx 
\frac{2\pi}{\beta}\left[-V^{\prime\prime}(\psi_{\max})V^{\prime\prime}(\psi_{\min})\right]^{-\frac{1}{2}}e^{\beta \Delta V},& \notag \\
\beta\rightarrow\infty,~f\rightarrow 0.&
\label{eq:0324_C_0928}
\end{align}

Finally, $C=\langle I_{\pm}(\psi)\rangle_{\psi}$ is derived as follows:
\begin{align}
C&= C\int_0^{2\pi}P_{S}(\psi)d \psi \notag \\
&=\int_0^{2\pi}\int_{\psi}^{\psi + 2\pi} e^{-\beta V(\psi)}e^{\beta V(\psi^{\prime})} d \psi^{\prime} d \psi \notag \\
&=\int_0^{2 \pi} \int_0^{2 \pi} e^{-\beta V(\psi)} e^{\beta V(\psi + y)} d \psi d y \notag \\
&=\langle I_{-}(\psi) \rangle_{\psi}~\left(=\langle I_{+}(\psi) \rangle_{\psi}\right),
\end{align}
where $\psi^{\prime}=\psi + y$ and $\frac{\partial (\psi, \psi^{\prime})}{\partial (\psi, y)}=
\begin{vmatrix}
1	&	0	\\
1	&	1	\\
\end{vmatrix}
=1$.\\
\

\section{\label{sec:appendix_E}Derivation of Eq.~(\ref{eq:0523_P_S(PSI)}) and Eq.~(\ref{eq:0725_tild_beta})}

By \cite[Chap. 2, Sec. 1, Theorem 1, Eq. (1.12)]{R.Wong1989}, the asymptotic approximation of the following integral
\begin{align}
\int_{\alpha}^{\gamma}\varphi(\psi)e^{\beta V(\psi)}d\psi
\approx e^{\beta V(\alpha)}\sum_{s=0}^{\infty}\Gamma\left(s+1\right)\beta^{-s-1}c_{s},& \notag \\
\hspace{2cm}\beta\rightarrow\infty,~f\rightarrow\infty&
\end{align}
is obtained under the conditions (i)-(iv) in \cite[p.~58]{R.Wong1989}, which are assumed in our setting for Eq.~(\ref{form_Ps}).
The coefficients $c_{s}$ are expressed in terms of $a_{s}$ and $b_{s}$ in the following manner.

First, as for $a_{s}$ and $b_{s}$, from $-V(\psi)\approx -V(\alpha) + \sum_{s=0}^{\infty}a_{s}(\psi - \alpha)^s$ and $\varphi(\psi)\approx\sum_{s=0}^{\infty}b_{s}(\psi - \alpha)^s$, by noting that $V(\psi)$ has the maximum at $\psi=\alpha$, we obtain the following equations:
\begin{align}
&-a_{0}=V^{\prime}(\alpha),
~-a_{1}=\frac{1}{2}V^{\prime\prime}(\alpha),
~a_{2}=\frac{1}{6}V^{\prime\prime\prime}(\alpha), \notag \\
&b_{0}=1,
~b_{n}=0~(n\ge1), \notag \\
&(\because\text{the above}~\varphi(\psi) = 1~\text{in Eq.~(\ref{form_Ps})}).
\end{align}
Then, by \cite[p.~58]{R.Wong1989}, we obtain $c_{s}$ as follows:
\begin{align}
c_{0}&= b_{0}/a_{0}=-V^{\prime}(\alpha)^{-1}, \notag \\
c_{1}&= -\frac{2a_{1}b_{0}}{a_{0}}\cdot\frac{1}{a_{0}^{2}}=-V^{\prime\prime}(\alpha)V^{\prime}(\alpha)^{-3}, \notag \\
c_{2}&= \left(4a_{1}^{2}-2a_{0}a_{2}\right)\cdot\frac{3}{2a_{0}^{2}}\cdot\frac{1}{a_{0}^{3}} \notag \\
&=-\frac{1}{2}\left[3V^{\prime\prime}(\alpha)^2V^{\prime}(\alpha)^{-5}-V^{\prime\prime\prime}(\alpha)V^{\prime}(\alpha)^{-4}\right].
\end{align}

\begin{figure}[h!]
\begin{minipage}{1\linewidth}
\begin{center}
\includegraphics[scale=0.4]{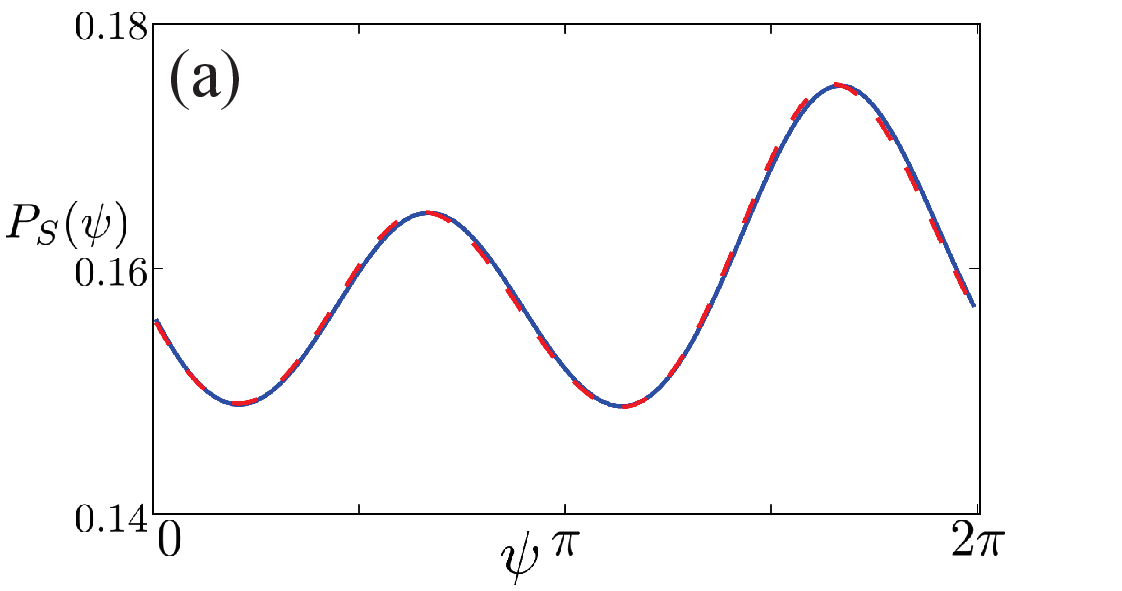}
\label{fig0523_a1}
\end{center}
\end{minipage}\\
\hspace{0.05\linewidth}
\begin{minipage}[b]{1\linewidth}
\begin{center}
\includegraphics[scale=0.4]{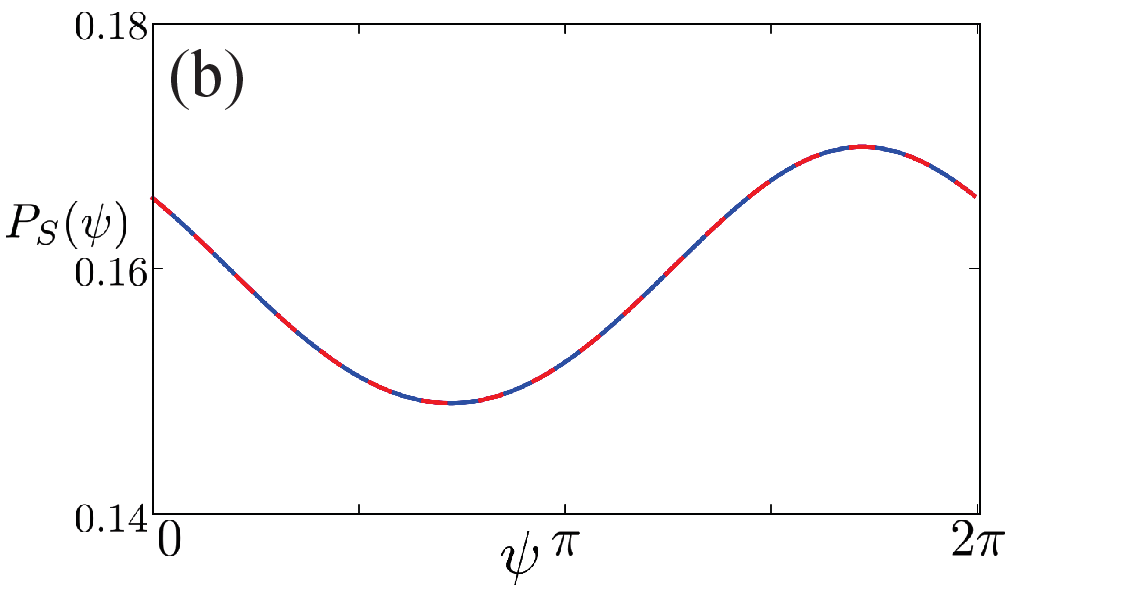}
\label{fig0523_1b}
\end{center}
\end{minipage}
\caption{Comparison between $P_{S}(\psi)$ in Eq.~(\ref{form_Ps}) (blue solid line) and its approximated (\ref{approx_Wong}) (red dashed line; Eq.~(\ref{eq:0523_P_S(PSI)}) in the main text) in case of (a) $r=0.2$, $f=2.8~(>f_{\max}=0.2736)$, $\beta=2.0$, and (b) $r=1.0$, $f=2.8~(>f_{\max}=0.2599)$, $\beta=10.0$.}
\label{fig0523}
\end{figure}

Substituting the above $c_{s}$ in Eq.~(\ref{form_Ps}), and noting the Gamma function $\Gamma(1)=1$, $\Gamma(2)=1$, and $\Gamma(3)=2$, the following Eq.~(\ref{approx_Wong}) holds up to $s=2$, after replacing $\alpha$ (in $V(\alpha)$) with $\psi$, which amounts to Eq.~(\ref{eq:0523_P_S(PSI)}):
\begin{widetext}
\begin{align}
P_{S}(\psi)
&= C^{-1}e^{-\beta V(\psi)}\int_{\psi}^{\psi + 2\pi} e^{\beta V(\psi^{\prime})}d \psi^{\prime} 
\approx C^{-1}e^{-\beta V(\psi)}e^{\beta V(\psi)}\sum_{s=0}^{\infty}\Gamma\left(s+1\right)\beta^{-s-1}c_{s} \notag \\
&\approx C^{-1}\left\{-\beta^{-1}V^{\prime}(\psi)^{-1} - \beta^{-2}\frac{V^{\prime\prime}(\psi)}{V^{\prime}(\psi)^{3}} - \beta^{-3}\left[\frac{3V^{\prime\prime}(\psi)^{2}}{V^{\prime}(\psi)^{5}}-\frac{V^{\prime\prime\prime}(\psi)}{V^{\prime}(\psi)^{4}}\right]\right\},
\hspace{2.5cm}\beta\rightarrow\infty,~f \rightarrow\infty.
\label{approx_Wong}
\end{align}
\end{widetext}
For the Example in Section~\ref{sec:beta-and-Z}, the accuracy of the asymptotic expansion (\ref{approx_Wong}) is rather high even for a relatively small $\beta$, as depicted in Fig.~\ref{fig0523};
for larger $\beta$, the graphs of Eq.~(\ref{form_Ps}) and Eq.~(\ref{approx_Wong}) become indistinguishable.

By plugging in $\psi = \psi_{\max,\min}$ in Eq.~(\ref{approx_Wong}) and using $V^{\prime}(\psi_{\max,\min}) = - J_{S}P_{S}(\psi_{\max,\min})^{-1}$, for sufficiently large $\beta$, we have
\begin{align}
\mathcal{C}_{S} &\equiv P_{S}(\psi_{\max})/P_{S}(\psi_{\min}) \notag \\
&\approx [P_{S}(\psi_{\max})/P_{S}(\psi_{\min})]\cdot R_{\max}/R_{\min},
\label{eq:0703C_S}
\end{align}
where $R_{\max,\min}$ is defined as 
\begin{align}
R_{\ast}\equiv 1
&+\beta^{-1}\frac{V^{\prime\prime}(\psi_{\ast})}{V^{\prime}(\psi_{\ast})^{2}} \notag \\
&+\beta^{-2}\left\{\frac{3V^{\prime\prime}(\psi_{\ast})^{2}}{V^{\prime}(\psi_{\ast})^{4}}-\frac{V^{\prime\prime\prime}(\psi_{\ast})}{V^{\prime}(\psi_{\ast})^{3}}\right\},
\label{eq:0710V_ast}
\end{align}
with the suffix $\ast$ representing $\max$ or $\min$.
The relationship in Eq.~(\ref{eq:0703C_S}) yields an asymptotic equality $R_{\max} \approx R_{\min}$ for sufficiently large $\beta$, which gives the following estimate $\beta_{A}$ of $\beta$: 
\begin{align}
\beta_{A}
=\left(C_{\min,2}-C_{\max,2}\right)/\left(C_{\max,1}-C_{\min,1}\right),
\label{eq:0704_tild_beta}
\end{align}
where
\begin{align}
C_{\ast,1} &= V^{\prime\prime}(\psi_{\ast})V^{\prime}(\psi_{\ast})^{-2},
~\text{and} \notag \\
C_{\ast,2} &= 3V^{\prime\prime}(\psi_{\ast})^{2}V^{\prime}(\psi_{\ast})^{-4}-V^{\prime\prime\prime}(\psi_{\ast})V^{\prime}(\psi_{\ast})^{-3}
\label{C_ast}
\end{align}
with the suffix $\ast$ representing $\max$ or $\min$.

\section{\label{sec:appendix_F} Sensitivity of inferred noise intensity}

In conjunction with \textit{Result 2} in the main text, we consider the following three different estimates Eq.~\eqref{beta_B}, Eq.~\eqref{widetilde_C_equiv}, and Eq.~(\ref{widetilde_C_equiv^ast}) for inferring $\beta$ only from the measured $P_S$.
First, after a simple (but lengthy) calculation, Eq.~(\ref{eq:0523_P_S(PSI)}) and the following equalities:
\begin{align}
&V^{\prime}(\psi_{\ast})
=-J_{S}P_{S}(\psi_{\ast})^{-1}, \notag \\
&V^{\prime\prime}(\psi_{\ast})
=-\beta^{-1}P_{S}(\psi_{\ast})^{-1}P_{S}^{\prime\prime}(\psi_{\ast})~\text{, and} \notag\\ 
&V^{\prime\prime\prime}(\psi_{\ast})
=J_{S}{P_{S}^{\prime\prime}(\psi_{\ast})}/{P_{S}(\psi_{\ast})^{2}}-\beta^{-1}{P_{S}^{\prime\prime\prime}(\psi_{\ast})}/{P_{S}(\psi_{\ast})},
\label{Vd_Vdd_Vddd}
\end{align}
with $\psi_{\ast}=\psi_{\max,\min}$, yield
\begin{align} 
J_{S}\beta~(\equiv \alpha) \approx D_{0}/D_{1}
\label{eq:0928_J_S}
\end{align}
for sufficiently large $\beta$, where 
\begin{align*}
D_{0}&=3 \left[P_{S}(\psi_{\max})^{2}P_{S}^{\prime\prime}(\psi_{\max})^2 -P_{S}(\psi_{\min})^2P_{S}^{\prime\prime}(\psi_{\min})^2\right], \\
D_{1}&= P_{S}(\psi_{\max})^{2}P_{S}^{\prime\prime\prime}(\psi_{\max})-P_{S}(\psi_{\min})^{2}P_{S}^{\prime\prime\prime}(\psi_{\min}).
\end{align*}
Thus, the following estimate of $\beta$ is obtained:
\begin{align}
\beta_B \equiv J_{S}^{-1}D_{0}/D_{1},
\label{beta_B}
\end{align}
which involves only the following observables:
$J_{S}$, $P_{S}(\psi_{\max,\min})$, $P_{S}^{\prime\prime}(\psi_{\max,\min})$, and $P_{S}^{\prime\prime\prime}(\psi_{\max,\min})$.

However, $\beta_B$ is no more accurate compared to $\beta_A$ obtained from $V$.
For instance, the Example (Eqs.~(\ref{Z_theta}) and (\ref{q_theta})) with $\beta=40.0$, $f=2.8$, and $J_S \simeq 0.444$, gives $\beta_B \simeq -2.160$.
The reason for this inaccuracy is explained by the following equation obtained from Eq.~(\ref{-Vd}):
\begin{align}
V(\psi)=-\int_0^{\psi}d\psi^{\prime}\left[J_{S} + \beta^{-1}P_{S}^{\prime}(\psi^{\prime})\right]P_{S}(\psi^{\prime})^{-1} ,
\label{V_psi}
\end{align}
which implies $V$ embeds the information of $\beta$ and $P_{S}$.
Namely, the RHS of Eq.~(\ref{eq:0725_tild_beta}) carries the information of the true $\beta$ (and $P_{S}$).
In contrast, $D_0$ and $D_1$ above carry only the information of $P_{S}$ and have less information of $\beta$.
The above explanation becomes more precise as presented with the other estimates of $\beta_C$ below in Eqs.~\eqref{widetilde_C_equiv} and \eqref{widetilde_C_equiv^ast}.

Next, we present yet another estimate of the noise intensity ($\beta_{C}$ in Eq.~(\ref{widetilde_C_equiv})) and its estimation accuracy is explained.
Going back to Eq.~(\ref{eq:0523_P_S(PSI)}) and using Eq.~(\ref{Vd_Vdd_Vddd}), Eq.~(\ref{eq:0523_P_S(PSI)}) is rewritten as
\begin{align}
C\approx\alpha^{-1}
&-\alpha^{-4}P_{S}(\psi_{\ast})^{2}P_{S}^{\prime\prime\prime}(\psi_{\ast}) \notag \\
&+3\alpha^{-5}P_{S}(\psi_{\ast})^{2}P_{S}^{\prime\prime}(\psi_{\ast})^{2}
\label{approxC_RHS}
\end{align}
with $\alpha=J_{S}\beta$ and $\psi_{\ast}=\psi_{\max,\min}$.
On the other hand, Eq.~(\ref{eq:0523_P_S(PSI)}) and Eq.~(\ref{Vd_Vdd_Vddd}) also yield the following asymptotic equation
\begin{align}
C&=C \int_{0}^{2\pi}P_{S}(\psi)d\psi
\approx\alpha^{-1}
+\alpha^{-4}\int_{0}^{2\pi}C_{4}(\psi)d\psi\notag\\ 
&+\alpha^{-5}\int_{0}^{2\pi}C_{5}(\psi)d\psi,
\label{approxC_LHS}
\end{align}
in which $C_{4}(\psi)$ and $C_{5}(\psi)$ are explicitly given in terms of $P_{S}(\psi)$, $P_{S}^{\prime}(\psi)$, $P_{S}^{\prime\prime}(\psi)$ and $P_{S}^{\prime\prime\prime}(\psi)$ in Eq.~(\ref{power_of_alpha}) at the end of this Appendix.
Now, in Eq.~(\ref{approxC_RHS}), if true (accurate) values of $C$ and $\psi_{\ast}~(=\psi_{\max,\min})$ are given, then by solving Eq.~(\ref{approxC_RHS}) for $\alpha$ ($\equiv J_{S}\beta$) with a true value of $J_{S}$, an estimate similar to $\beta_B$ is given as
\begin{align}
\beta_C\equiv J_{S}^{-1}\alpha,
\label{widetilde_C_equiv}
\end{align}
which is found to be quite accurate, with an estimation error of $<0.5\%$ (for the Example with $f=2.8$ and any $\beta>2.0$).

In contrast, if the $C$ in Eq.~(\ref{approxC_RHS}) is replaced with the approximated value by the asymptotic (\ref{approxC_LHS}), the resulting estimate
\begin{align}
\beta_C^{\ast}=\widetilde{J_{S}}^{-1}\alpha
\label{widetilde_C_equiv^ast}
\end{align}
with $\alpha$ being obtained from Eq.~(\ref{approxC_RHS}) (for the observed $J_{S}$) is explicitly expressed in terms of the observables $\widetilde{J_{S}}$, $\widetilde{P_{S}}(\psi_{\ast})$, $\widetilde{P_{S}^{\prime}}(\psi_{\ast})$, $\widetilde{P_{S}^{\prime\prime}}(\psi_{\ast})$ and $\widetilde{P_{S}^{\prime\prime\prime}}(\psi_{\ast})$, but this ${\beta_C^{\ast}}$ turns out to be no more accurate compared to the ${\beta_C}$ in Eq.~(\ref{widetilde_C_equiv}); 
for instance, the Example with $\beta=5.0$, $f=2.8$, and $\widetilde{J_{S}}\simeq0.4429$ gives $\beta_C^{\ast}\simeq2.879$.
This is because in Eq.~(\ref{approxC_RHS}) both of the $O\left(\alpha^{-2}\right)$ and $O\left(\alpha^{-3}\right)$ terms vanish, which results in the situation where a `small' error of $C$ from the approximation (\ref{approxC_LHS}) is significantly magnified in ${\beta_C^{\ast}}$ with respect to the accurate ${\beta_C}$.

This mechanism for the magnified error (due to the absence of the $O\left(\alpha^{-2}\right)$ and $O\left(\alpha^{-3}\right)$ terms in Eq.~(\ref{approxC_RHS})) accounts for the inaccuracy of ${\beta_C^{\ast}}$ above, which can be explained in detail in the following manner:

First, for $\beta\gg1$, from Eq.~(\ref{form_Js}) $\alpha C = J_{S}\beta C = 1-e^{-2\pi f\beta} \approx 1$.

Second, from either Eq.~(\ref{approxC_RHS}) and Eq.~(\ref{approxC_LHS}), $\alpha C = 1 + O\left(\alpha^{-3}\right)\approx1~(\alpha\gg1)$, which is consistent with the above $\alpha C \approx 1$.

Finally, $J_{S} \approx$ constant ($\beta\gg1$) is verified in the following manner:\\
(i) From the definition of $J_{S}$, $J_{S}=[d+H(\psi)]P_{S}(\psi)-\beta^{-1}P_{S}^{\prime}(\psi)$.\\
(ii) For the rotating phase (for $f>f_{\max}$), $P_{S}(\psi)\sim(\dot{\psi})^{-1}$ for $\beta\gg1$ with $\dot{\psi}=d+H(\psi)$ (cf. \cite[p.~282]{Pikovsky2001}, \cite[p.~73]{Kuramoto}).\\
From (i) and (ii), and noting that $P^{\prime}_{S}(\psi_{\ast})=0$ for $\psi_{\ast}=\psi_{\max,\min}$, $J_{S}=[d+H(\psi_{\ast})]C_{\ast}[d+H(\psi_{\ast})]^{-1}=C_{\ast}$ with $C_{\ast}\equiv\left\{\int_{0}^{2\pi}[d+H(\psi)]^{-1}d\psi\right\}^{-1}$, which implies $J_{S}$ becomes a constant independent from $\beta$ as $\beta\rightarrow\infty$.

Thus, for $\beta\gg1$, the inferred $\beta$ is sensitive to a small variation of $C$ since $\beta \approx J_{S}^{-1}C^{-1}$ with $C\ll1$.

Finally, $C_4(\psi)$ and $C_5(\psi)$ are given in the following manner:
substituting Eq.~(\ref{Vd_Vdd_Vddd}) in Eq.~(\ref{eq:0523_P_S(PSI)}) yields Eq.~(\ref{approxC_LHS}), and abbreviating $P_S(\psi)$, $P_S^{\prime}(\psi)$, $P_S^{\prime\prime}(\psi)$, and $P_S^{\prime\prime\prime}(\psi)$ as $P_S$, $P_S^{\prime}$, $P_S^{\prime\prime}$, and $P_S^{\prime\prime\prime}$, respectively, we obtain
\begin{align*}
C_4(\psi)&= -P_{S}P_{S}^{\prime3}-4P_{S}^{2}P_{S}^{\prime}P_{S}^{\prime\prime}-P_{S}^{3}P_{S}^{\prime\prime\prime}, \\
C_5(\psi)&= 3P_{S}P_{S}^{\prime4}+16P_{S}^{2}P_{S}^{\prime2}P_{S}^{\prime\prime}+3P_{S}^{3}P_{S}^{\prime\prime2}+4P_{S}^{3}P_{S}^{\prime}P_{S}^{\prime\prime\prime}.
\end{align*}
Their derivation is as given below, to the power of $\alpha~(=J_{S}\beta)$:
\begin{widetext}
\begin{align}
&C\int_{0}^{2\pi}P_{S}(\psi)d\psi
\approx\int_{0}^{2\pi}\left\{-\beta^{-1}V^{\prime}(\psi)^{-1} - \beta^{-2}\frac{V^{\prime\prime}(\psi)}{V^{\prime}(\psi)^{3}} - \beta^{-3}\left[\frac{3V^{\prime\prime}(\psi)^{2}}{V^{\prime}(\psi)^{5}}-\frac{V^{\prime\prime\prime}(\psi)}{V^{\prime}(\psi)^{4}}\right]\right\}d\psi \notag \\
&=\alpha^{-1}+0+0 \notag \\
&+\alpha^{-4}\int_{0}^{2\pi}\left[
P_{S}\left(2P_{S}^{\prime3}+3P_{S}P_{S}^{\prime}P_{S}^{\prime\prime}\right)-P_{S}\left(3P_{S}^{\prime3}+7P_{S}P_{S}^{\prime}P_{S}^{\prime\prime}+P_{S}^{2}P_{S}^{\prime\prime\prime}\right)
\right]d\psi \notag \\
&+\alpha^{-5}\int_{0}^{2\pi}\left[
-P_{S}\left(3P_{S}^{\prime4}+6P_{S}P_{S}^{\prime2}P_{S}^{\prime\prime}\right)+P_{S}\left(6P_{S}^{\prime4}+22P_{S}P_{S}^{\prime2}P_{S}^{\prime\prime}+3P_{S}^{2}P_{S}^{\prime\prime2}+4P_{S}^{2}P_{S}^{\prime}P_{S}^{\prime\prime}\right)
\right]d\psi \notag \\
&=\alpha^{-1}
+\alpha^{-4}\int_{0}^{2\pi}\left(
-P_{S}P_{S}^{\prime3}-4P_{S}^{2}P_{S}^{\prime}P_{S}^{\prime\prime}-P_{S}^{3}P_{S}^{\prime\prime\prime}
\right)d\psi
+\alpha^{-5}\int_{0}^{2\pi}\left(
3P_{S}P_{S}^{\prime4}+16P_{S}^{2}P_{S}^{\prime2}P_{S}^{\prime\prime}+3P_{S}^{3}P_{S}^{\prime\prime2}+4P_{S}^{3}P_{S}^{\prime}P_{S}^{\prime\prime\prime}
\right)d\psi. \notag \\
&\equiv
\alpha^{-1} + \alpha^{-4}\int_0^{2\pi}C_{4}(\psi) d\psi + \alpha^{-5}\int_0^{2\pi}C_{5}(\psi)d\psi.
\label{power_of_alpha}
\end{align}
\end{widetext}

\section{\label{sec:appendix_G}Accuracy of various quantities}

The following tables show the estimated values and their errors.
For all quantities, the errors are less than $5\%$, showing reasonable accuracy of our framework.
\begin{table}[htpb] 
\caption{Measured observables $\left(\widetilde{f}, \widetilde{\mathcal{C}_{S}}, \widetilde{\Delta\psi}\right)$ (from SDE (\ref{eq:0428_dpsi/dtau}) with $\beta=10.0$);
``error'' indicates the estimation error for the corresponding true value of ($f$, $\mathcal{C}_{S}$, $\Delta\psi$).
(a) The case of $r=0.2$, $f_1=0.28\times10^{-2}$, and $f_2=0.26\times10^{-1}$;
(b) the case of $r=1.0$, $f_1=0.2\times10^{-2}$, and $f_2=0.24\times10^{-2}$.
}\label{table_experiment2-1}
\vspace{0.1cm}
\begin{tabular}{c|c|c}
\hline
(a)~~~$\widetilde{f}$/error~~~~~~~ &	$\widetilde{\mathcal{C}_{S}}$/error &	$\widetilde{\Delta \psi}$/error \rule[0pt]{0pt}{10pt}\\
\hline\hline
0.271$\times 10^{-2}$/3.15$\%$ &	22.6236/4.11$\%$ &	4.4234/0.37$\%$ \\
\hline
0.252$\times10^{-1}$/3.08$\%$ &	22.7212/4.19$\%$ &	4.3982/0.57$\%$ \\
\hline
\end{tabular}
\\
\vspace{0.1cm}
\vspace{0.5cm}
\begin{tabular}{c|c|c}
\hline
(b)~~~$\widetilde{f}$/error~~~~~~~ &	$\widetilde{\mathcal{C}_{S}}$/error &	$\widetilde{\Delta \psi}$/error \rule[0pt]{0pt}{10pt} \\
\hline\hline
0.195$\times 10^{-2}$/2.64$\%$ &	38.5340/1.66$\%$ &	3.1165/0.40$\%$ \\
\hline
0.234$\times 10^{-2}$/2.38$\%$ &	38.5055/1.71$\%$ &	3.1353/0.28$\%$ \\
\hline
\end{tabular}
\end{table}

\begin{widetext}
\begin{table}[htpb] 
\caption{Accuracy of $\beta_{A}$ for various $\beta$;
\textit{EE} indicates the estimation error $(=(\beta - \beta_{A})/\beta)$ defined in Numerical verification for Result 2.
}
\label{tab:estimate_beta=2}
\vspace{0.2cm}
\centering
\begin{tabular}{c|c|c|c|c|c|c}
\hline 
$\beta$	&	$\beta_A$  &	\textit{EE} [$\%$] &	$C_{\max,1}$ &	$C_{\min,1}$ &	$C_{\max,2}$ &	$C_{\min,2}$ \rule[0pt]{0pt}{10pt}	\\
\hline\hline
10	&	9.89 &	1.134 &	$-2.822\times10^{-3}$ &	$5.294\times10^{-3}$ &	$2.787\times10^{-2}$ &	$-5.237\times10^{-2}$  \\
\hline
20 &	19.93 &	0.373	&	$-1.410\times10^{-3}$ &	$2.636\times10^{-3}$ &	$2.804\times10^{-2}$ &	$-5.257\times10^{-2}$  \\
\hline
30 &	29.94 &	0.206	&	$-9.401\times10^{-4}$ &	$1.756\times10^{-3}$ &	$2.810\times10^{-2}$ &	$-5.261\times10^{-2}$  \\
\hline
40 &	39.94 &	0.138	&	$-7.051\times10^{-4}$ &	$1.316\times10^{-3}$ &	$2.812\times10^{-2}$ &	$-5.262\times10^{-2}$  \\
\hline
50	 &	49.95 &	0.103 &	$-5.640\times10^{-4}$ &	$1.053\times10^{-3}$ &	$2.814\times10^{-2}$ &	$-5.263\times10^{-2}$ \\
\hline
75 &	74.95 &	0.062	& $-3.760\times10^{-4}$ & $7.019\times10^{-4}$ &	$2.816\times10^{-2}$ &	$-5.263\times10^{-2}$  \\
\hline
100 &	99.96 &	0.044 &	$-2.820\times10^{-4}$ &	$5.264\times10^{-4}$ &	$2.817\times10^{-2}$ &	$-5.264\times10^{-4}$ \\
\hline
\end{tabular}
\end{table}
\end{widetext}

\begin{widetext}
\begin{table}[H] 
\caption{Fourier coefficients of reconstructed $Z$ from Eqs.~(\ref{slope}) and (\ref{eq:0523_alpha_nbeta_n}).
(a) The case of a doubly-humped $Z$ with $r=0.2$;
(b) the case of a single-peaked $Z$ with $r=1.0$.
}
\label{table_0612_Fourier_coefficients}
\vspace{0.2cm}
\begin{tabular}{c|c|c|c|c|c}
\hline
(a) & $Z_0$ & $c_1$ & $c_2$ & $d_1$ & $d_2$ \\
\hline\hline
true value & 0.0 & 0.8944 & 1.7889 & 0.0	&	0.0 \\
\hline
estimated value	&	$-0.4053$	&	0.9119	&	1.7820	&	0.0101	&	$9.5220\times10^{-4}$	\\
\hline
\end{tabular}
\\
\vspace{0.2cm}
\centering
\begin{tabular}{c|c|c|c|c|c}
\hline
 (b) & $Z_0$ & $c_1$ & $c_2$ & $d_1$ & $d_2$ \\
\hline\hline
true value & 0.0 & 2.0 & 0.0 & 0.0 & 0.0 \\
\hline
estimated value	&	$-0.0347$ &	2.0692 &	$-9.6774\times10^{-4}$ &	$-0.0047$	&	0.0082 \\
\hline
\end{tabular}
\end{table}
\end{widetext}

\section{\label{sec:appendix_H}Approximation error of Eq. \eqref{eq:0515_C_S}}

The approximation error ($AE$) of $\mathcal{C}_S$ in Eq.~(\ref{eq:0515_C_S}) grows up as $\vert f \vert$ increases from $0$.
For $r=0.2$,
(i) with $f=0.28\times 10^{-3}$, all \textit{AE}s $<1\%$ for $\beta=2.5$, $5.0$, $7.5$, $10$ and $20$,
(ii) with $f=0.28\times 10^{-2}$, all \textit{AE}s $<3\%$, for all $\beta$ above, and
(iii) with $f=0.28\times 10^{-1}$, all \textit{AE}s $<6\%$ for $\beta=2.5$, $5.0$ and $7.5$.
Note that, in all these (successful) cases, $\psi_{\max,\min} \simeq \psi_{u,s}$ holds.
The reason for this increasing error of $\mathcal{C}_S$ is, however, due to a small error between $\psi_{\max,\min}$ and $\psi_{u,s}$, which grows  exponentially in the term $e^{\beta \Delta V}= e^{\beta(V_{\min}-V_{\max} + f \Delta \psi)}$ of Eq.~(\ref{eq:0515_C_S}) for larger $\beta$ as $\vert f \vert$ increases because the measured $\widetilde{\Delta \psi}= \psi_{\max} - \psi_{\min}~ \left( \approx \psi_{s} - \psi_{u} \equiv \Delta \psi \right)$ involves a small error of $\widetilde{\Delta \psi} - \Delta \psi$.

\section{\label{sec:appendix_I}Inference of noise intensity from only one set of observables}
Instead of using the two sets of the observable $\mathcal{C}_{S,i}$ in Eq.~(\ref{C_S,i}), if only one set of the observables $P_{S}(\psi)$ and $J_{S}$ is available, the unknown $\beta$ and $\Delta V$ can be obtained as follows.

First, the following asymptotic approximations hold for sufficiently large $\beta$:
\begin{subequations}
\begin{align}
P_{S}(\psi_{\max})
& \approx (A/C) e^{\beta \Delta V - \beta 2 \pi f}
= B e^{-\beta 2 \pi f},
\label{Ps_psi_max} \\
P_{S}(\psi_{\min})
& \approx \frac{A\left(1+e^{-\beta 2\pi f}\right)}{2C}
= \frac{B\left(1+e^{-\beta2\pi f}\right)}{2e^{\beta\Delta V}},
\label{Ps_psi_min}
\end{align}\label{Ps_psi_max_min}
\end{subequations}
with
\begin{align}
A &\equiv \sqrt{(2\pi/\beta)[-V^{\prime\prime}(\psi_{\min})]^{-1}}, \notag \\
B &\equiv \sqrt{(\beta/2\pi)V^{\prime\prime}(\psi_{\max})} ,~\text{and} \notag \\
C &\approx (2\pi/\beta)[-V^{\prime\prime}(\psi_{\max})V^{\prime\prime}(\psi_{\min})]^{-\frac{1}{2}}e^{\beta\Delta V},~\text{with} \notag \\
\Delta V &= V_{\min} - V_{\max} + f\Delta\psi,
\label{ABCDeltaV}
\end{align}
which are directly obtained from Laplace's formula and are consistent with Eq.~(\ref{eq:0515_C_S}).

Next, $V^{\prime\prime}(\psi_{\max,\min})=-\beta^{-1}P_S(\psi_{\max,\min})^{-1}P_S^{\prime\prime}(\psi_{\max,\min})$ is obtained immediately by differentiating both sides of Eq.~(\ref{-Vd}).
Finally, $f$ is obtained from Eq.~(\ref{define_Q}) with the measured  $J_{S}$ and $P_{S}(\psi)$.
Thus, we obtain all the measured elements in Eqs.~\eqref{define_Q} and \eqref{-Vd} for inferring $\beta$ and $\Delta V$.
For the derivation of Eqs.~(\ref{Ps_psi_max_min}) and (\ref{ABCDeltaV}), see Appendix~\ref{sec:appendix_A} and Appendix~\ref{sec:appendix_D};
they are obtained as in Eqs.~(\ref{eq:1208_P_S_psi_min}), (\ref{eq:1208_P_S_psi_max}), and (\ref{eq:0324_C_0928}), respectively.

Although Eqs.~(\ref{Ps_psi_max},~b) are simpler than Eq.~(\ref{C_S,i}), we verified that for the example with $r=0.2$, the approximated equations can adequately be used for sufficiently large $\beta$ (e.g., $\beta>50$).
For small $\beta$ ($\beta<50$), the approximation error for $C$ is not negligible, because the above approximation of $C$ is obtained from Laplace's formula.
This situation is in contrast to the one in Eqs.~(\ref{eq:C_S_1003}), (\ref{eq:0515_C_S}) and (\ref{C_S,i}), in which the normalization constant $C$ is conveniently eliminated.

\section{\label{sec:appendix_J}Experimentally determined coupling function}
The pioneering study in \cite{Miyazaki2006} has obtained the coupling function $H(\psi)$ for two mutually coupled chemical oscillators, which is virtually identified with an externally forced single oscillator in Eq.~(\ref{eq:0428_dpsi/dtau}) if we introduce the phase difference $\psi$ between the two phases of the two oscillators.
We note that the same argument holds true for the mutually coupled noisy oscillators (cf. \cite[pp. 245--246]{Pikovsky2001}).
The method in \cite{Miyazaki2006} is simple and practical since it measures only the return time $T+\Delta T$, beginning from each oscillation phase $\psi$ and then going back to the same phase, with $T$ being the natural period of the oscillator and $\Delta T~(=\Delta T(\psi))$ being a periodic function of $\psi$;
this $\Delta T(\psi)$ directly constructs the coupling function $H(\psi)$.
The experimental result in \cite[Fig.~3]{Miyazaki2006} presents (slightly noisy but) consistent data for three different coupling strengths for the case of the rotating phase ($\dot{\psi}>0$; $f>f_{\max}$ in our setting).

\newpage
\nocite{Tiana-Alsina2005}

\bibliography{PRResearch2023_bib_29}
\end{document}